\pdfoutput=1
\documentclass[12pt]{article}
\usepackage{graphicx}
\usepackage{amsmath,amssymb}
\usepackage{pdfsync}
\def\a{\alpha}

\def\d{\delta}
\def\D{\Delta}
\def\F{\Phi}
\def\f{\phi}
\def\r{\rho}

\def\s{\sigma}
\def\S{\Sigma}
\def\o{\omega}
\def\O{\Omega}
\def\p{\psi}

\def\m{\mu}
\def\n{\nu}
\def\t{\tau}
\def\x{\xi}

\def\nn{\nonumber}

\begin{document}

\thispagestyle{empty}
\setcounter{page}{0}
\renewcommand{\theequation}{\thesection.\arabic{equation}}

{\hfill{ULB-TH/10-35}}

\vspace{1 cm}

\begin{center} {\bf \Large The hidden horizon and black hole unitarity}

\vspace{.3 cm}

Fran\c {c}ois Englert${}^{a}$ and Philippe Spindel${}^{b}$

\footnotesize
\vspace{.4 cm}

${}^a${\em Service de Physique Th\'eorique, Universit\'e Libre de
Bruxelles\\Campus Plaine C.P.225, Boulevard du Triomphe, B-1050
Bruxelles, 
Belgium\\and\\ The International Solvay Institutes \\Campus Plaine
C.P.231, Boulevard du Triomphe, B-1050 Bruxelles, 
Belgium}

\vspace{.2cm}

${}^b${\em Service de M\'ecanique et Gravitation,
Universit\'e de Mons\\ Facult\'e des Sciences, 20, Place du Parc, B-7000 Mons,
Belgium}

 {\tt
fenglert@ulb.ac.be , philippe.spindel@umons.ac.be
} 

\end{center}

\vspace {.5 cm}

\begin{center}
{\bf Abstract} 

\begin{tabular}{p{15cm}}
{\small
\noindent

We motivate through a detailed analysis of the Hawking radiation in a Schwarzschild background a  scheme in accordance with quantum unitarity. In this scheme the semi-classical approximation of the unitary quantum -- horizonless -- black hole S-matrix leads to the conventional description of the Hawking radiation from a classical black hole endowed with an event horizon. Unitarity is borne out by the detailed {\it exclusive} S-matrix amplitudes. There, the fixing of generic out-states, in addition to the in-state, yields in asymptotic Minkowski space-time saddle-point contributions which are dominated by Planckian metric fluctuations when approaching the Schwarzschild radius. We argue that these prevent the corresponding macroscopic ``exclusive backgrounds'' to develop an event horizon.  However, if no out-state is selected,  a distinct saddle-point geometry can be defined, in which Planckian fluctuations are tamed. Such ``inclusive background'' presents an event horizon and constitutes a coarse-grained average over the aforementioned exclusive ones.  The classical event horizon  appears as a coarse-grained structure, sustaining the thermodynamic significance of the Bekenstein-Hawking entropy. This is reminiscent of the tentative fuzzball description of extremal black holes: the role of microstates is played here by a complete set of out-states. Although the computations of unitary amplitudes would require a detailed theory of quantum gravity, the proposed scheme itself, which appeals to the metric description of gravity only in the vicinity of stationary points, does not.}

\end{tabular}
\end{center}

\newpage
\tableofcontents
\newpage

\section{Introduction}
In the ``conventional" description of black hole evaporation~\cite{Hawking:1974rv} correlations between the structureless Hawking thermal radiation and the information contained in the space-time beyond the black hole horizon disappear in the singularity, leading to a violation of unitarity~\cite{Hawking:1976ra}, except if the black hole terminates on a problematic infinitely degenerate remnant~\cite{Susskind:1995da}.

Disregarding this possibility, one seems to be faced with an alternative: either there is, as in the original Hawking derivation, no information in the thermal state and unitarity is violated, or the original information is contained in the radiation and  unitarity is preserved, which leads to suspect that the original field theoretic derivation was essentially incorrect. Violating unitarity would have dramatic consequences for energy conservation~\cite{Banks:1983} and is at odd with string theories and with the AdS/CFT correspondence. On the other hand, the simplicity of the Hawking derivation and the consistency of its conclusions with the Bekenstein entropy~\cite{Bekenstein:1973} makes one reluctant to reject it. In the scheme proposed here, the Hawking description would emerge naturally from a unitary black hole S-matrix in a semi-classical approximation.

We first remark that in conventional relativistic field theory the S-matrix describes elementary particle interactions in a  fixed background, namely Minkowski space-time. If coupling to gravity is considered, different S-matrix elements may  require different corresponding backgrounds. We shall argue that in the presence of a  horizon this constitutes a dramatic effect~\cite{Englert:1994} which provides a clue for the solution of the black hole information paradox. More generally, this effect sheds light on the significance of the classical approximation of the quantum theory.

We discuss exclusive and inclusive S-matrix elements for Schwarzschild black holes.

Exclusive S-matrix amplitudes are scalar products $\langle f\vert i \rangle$ relating the Heisenberg in-state $\vert i\rangle$ of the initial constituents of a mass $M$ Schwarzschild black hole to a particular ``post-selected"~\cite{ABL: 1964, AC} final out-state $\vert f\rangle \neq \vert i\rangle$ of the decay products. We provide some evidence that, for suitable initial states $ \vert i\rangle$ and generic final states  $ \vert f\rangle$, the  saddle-points in the path integral describing $\langle f\vert i \rangle$  define ``exclusive backgrounds'' upon which Planckian quantum fluctuations prevent the occurrence of an event horizon.  This is taken as an indication that the black hole S-matrix is unitary, as originally proposed by 't Hooft~\cite{'tHooft:1984re, Stephens:1993an}. It was viewed as a complementarity structure by Susskind and al.~\cite{Susskind:1993}. The present interpretation of S-matrix unitarity is more directly related to the ``fuzzball'' conjecture~\cite{MST}.

Inclusive S-matrix amplitudes are defined by fixing only the initial state $\vert i\rangle$ generating a particular mass M black hole, thus avoiding post-selection. The black hole inclusive amplitude is a quantum superposition of its exclusive  amplitudes and, as such, it inherits a quantum superposition of its exclusive horizonless backgrounds, highly sensitive to Planckian effects.  However it is possible to select a unique alternate saddle-point yielding a geometry insensitive to Planckian metric fluctuations. This geometry is endowed with an event horizon.  We take it as defining the classical approximation to the quantum black hole.  We label the corresponding geometry the ``inclusive background''  and we show that it is an {\it average} over the exclusive backgrounds. The averaging means that the classical event horizon, and the classical world, is the result of  a coarse-graining of the full quantum description and has only thermodynamical significance. It is in this inclusive background that the Hawking radiation appears as a valid approximation.

These considerations are inferred from  a detailed analysis of the Hawking radiation in conventional field theory.  However, although we shall formally write functional integrals over metrics, we shall limit their use in the vicinity of stationary points. Thus the non-renormalizability of quantized  Einstein general relativity, as well as the problem of the measure and of the ghosts linked to the gauge invariance of the theory, and  even the possible emergence of the  metric from a more fundamental structure are not expected to alter significantly our considerations.

While in this scheme for solving the black hole paradox, the full computations of the exclusive amplitudes of the unitary S-matrix would require a well-defined detailed theory of quantum gravity, the scheme itself does not.

 In Section 2, we study in general terms the   back-reaction of quantum matter on gravity in the vicinity of  self-consistent backgrounds defined by saddle-points of the gravity-matter amplitudes.  For exclusive   amplitudes with out-state $\vert f \rangle$ selected, such exclusive backgrounds arise as stationary points of path integral evaluations.  These yield  (possibly complex) metrics driven by the ``weak value"~\cite{AC} of the energy-momentum tensor  operator $ \hat T_{\mu\nu}$ between the in- and out-states $\vert i \rangle$ and $\vert f \rangle$, namely $\langle f\vert \hat T_{\m\n}\vert i \rangle/\langle f\vert i\rangle$.  Leaving the out-state partially unspecified allows for alternate saddles which yield equations driven by generalized weak-values~\cite{MP, Brout:1995rd}. If the out-state is left totally unspecified,  the generalized weak value reduces to the usual mean value $\langle i\vert \hat T_{\m\n}\vert i\rangle$. The resulting inclusive background defines a real metric driven by the expectation value of the energy momentum tensor in the in-state. 

 In Section 3 we use these results to discuss the back-reaction of the Hawking radiation on the  unperturbed background metric depicted by the Penrose diagram of a classical  incipient black hole. In absence of post-selection, the inclusive background driven by the renormalized average energy-momentum tensor preserves the event horizon and allows for a semi-classical description. This is not the case if a generic out-state is selected on $\cal J^+$.  After reviewing in a simple model the standard derivation of the Hawking radiation and the general decomposition of the weak energy momentum tensor~\cite{MP,Englert:1994qe}, we examine the consequences of selecting a quantum state on $\cal J^+$. We show that generic post-selection induces huge perturbations of the geometry at a Planckian distance of the horizon, invalidating a semi-classical description of the back reaction. We take this as evidence that the classical horizon of an  unperturbed background cannot survive quantum fluctuations and that the aforementioned Penrose diagram is inconsistent for such post-selections. 

In Section 4 we argue that these inconsistencies persist in the full history of an incipient black hole and that post-selection yields, in a background void of classical horizon, an exclusive amplitude of a {\em unitary} S-matrix. We show how  the horizonless unitary S-matrix can hide a coarse-grained  horizon. We stress the implication of coarse-graining for the emergence of the classical world and its relation with the fuzzball conjecture.

\section{The back-reaction of quantum matter on gravity}

\subsection{The background for matter-gravity amplitudes}

Consider the amplitude for matter and gravity between a Heisenberg in-state $\vert i\rangle  $, defined by a functional  ${\bf \Psi_i}$ acting on a configuration $\chi_i\equiv \{\f_j^{(i)},\, g^{(i)}_{\m\n}\}$ of matter and gravity fields $\{\f_j,\, g_{\m\n}\}$ on a given spatial hypersurface $\S_0$, and a Heisenberg out-state $\vert f \rangle $, defined by the functional ${\bf \Psi _f} $ acting on configurations  $\chi_f\equiv \{\f_j^{(f)},\, g^{(f)}_{\m\n}\}$ specified  on a hypersurface $\S_1$. The amplitude $ \langle f\vert i \rangle $ is formally given by the path integral
\begin{equation}
\label{amplitude}
 \langle f \vert i \rangle  =\int{\cal D}\chi_f{\cal D}\chi_i  {\bf \Psi _f}^*[\chi_f]{\bf \Psi_i}[\chi_i] \prod_t{\cal D}(\{\f_j^{(t)}\}) {\cal D}(g_{\m\n}^{(t)})e^{i{\cal S} (\f_j, g_{\m\n};\chi_i,\chi_f)}\, ,
\end{equation}
where $\cal S$ is an action functional and the functionals  ${\bf \Psi _i}$ and ${\bf \Psi _f}$ are evaluated on the lower and upper limits of  the Green functional 
\begin{equation}
\label{Greenfunct}
U[\chi_i,\chi_f] =\int_{\chi_i}^{\chi_f} \prod_t{\cal D}(\{\f_j^{(t)}\}){\cal D}(g_{\m\n}^{(t)})e^{i{\cal S} (\f_j, g_{\m\n};\chi_i,\chi_f)}\, .
\end{equation}
Here what we call a field configuration on a given hypersurface is a set of classical values of the matter and gravitational fields (satisfying the required constraint equations). In other words the configuration on the hypersurface $\S_t$ denoted by $\chi_t$ is the equivalent of the position of a point particle, at time $t$, in classical  mechanic. Functional of configurations is a representative of a state.  It is the analogous of a wave function in  point particle quantum mechanic. The Green functional is formally given by an integration on all the field configurations that interpolate between the initial and final configurations $\chi_i$ and $\chi_f$, weighted by the exponential of $i$ times the action ${\cal S} (\f_j, g_{\m\n};\chi_i,\chi_f)$ evaluated on the interpolating fields. In the expression (\ref{amplitude}) of the amplitude $\langle f\vert i\rangle$ we didn't mention any more these boundary configurations because we have integrated on them by acting with the functional ${\bf \Psi_i}$ on $\chi_i$ and ${\bf \Psi_f}$ on $\chi_f$.

In order to avoid misunderstanding let us emphasize a conceptual point, on which all the forthcoming discussion will rest. 

We view metric as a possibly emergent concept but we nevertheless assume that  in the vicinity of saddle-points  of some hitherto unknown path integral a geometrical interpretation does exist.  This is in line with the interpretation of general relativity as a low energy approximation of a possibly more fundamental theory. Thus when we say that the path integral is ``formally'' given by  Eq.(\ref{amplitude})  we do not simply refer to the precise mathematical definition of such an object, with all the difficulties linked to the measure, the Faddeev-Popov ghosts, etc. We mean that
 Eqs.(\ref{amplitude}) stands for a much more complex expression which can be expressed in terms of genuine metric fields $g_{\m\n}$ only in the vicinity of saddle-points. Obviously, in what follows we shall not attempt to evaluate the deep quantum regime. We shall use the explicit form of Eqs.(\ref{amplitude}), restricted to its Gaussian approximation for the metric, keeping in mind that this procedure describes semi-classical configurations only if higher order quantum corrections are small. 

To make contact with the black hole space-time of a collapsing star,  we split the fields $\f_j$ into classical fields  $\F_j^c$ representing the star and genuine quantum fields describing the Hawking radiation in absence of back-reaction. To have a qualitative understanding of its back reaction, we shall take a single scalar quantum field $\f$. In the absence of back-reaction, the region outside the star is described by the Schwarzschild metric
\begin{equation}
ds^2= \left(1- \frac{2M}{r }\right) du\  dv - r^2
d\O^2\, , 
\label{Swartz}
\end{equation}
with
\begin{eqnarray}
u& =& t-r^* \qquad v=t+r^*\, , \label{uv}\\ 
dr& =& (1- \frac{2M}{r}) dr^*\, \label{tortoise}.  
\end{eqnarray}

\begin{figure}[h]
   \centering
   \includegraphics[width=6cm]{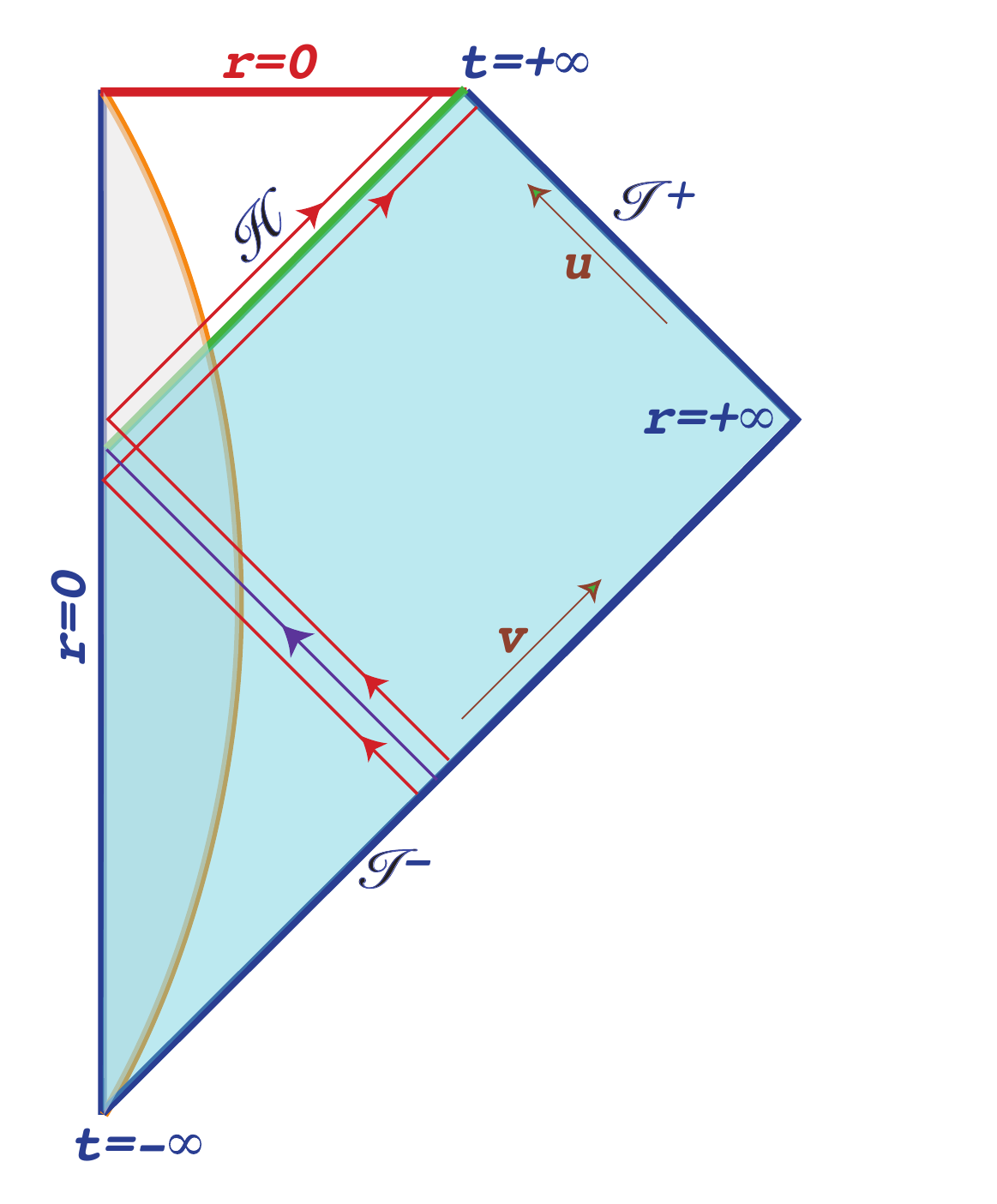} 
 \caption { \sl \small Penrose diagram of a collapsing shell. The horizon grows from 0 to 2M at the star surface and remains constant outside the star. The trajectories of an s-wave Hawking light ray and of its partner are depicted in red. The region outside the horizon is painted in blue and the Minkowskian region inside the star in grey.} 
\end{figure}

The Penrose diagram of the black hole metric is depicted in Fig.1, where the star is mimicked  by a shell. The tortoise coordinate $r^*$ spans the entire region outside the shell and outside the horizon. The advanced and retarded times $v$ and $u$ can be extended inside the star  by requiring continuity of the metric across the shell. 
 
Let ${\cal S}_g (g_{\m\n},\F_j^c)$ be the action in absence of the quantum field $\f$.   The amplitude Eq.(\ref{amplitude}) reads, with ${\cal D}\chi_i={\cal D}(\f^{(i)}){\cal D}(g^{(i)}_{\m\n})$,
${\cal D}\chi_f={\cal D}(\f^{(f)}){\cal D}(g^{(f)}_{\m\n})$, 
\begin{equation}
\label{field}
\langle f \vert i \rangle = \int{\cal D}\chi_f{\cal D}\chi_i  {\bf \Psi _f}^*[\chi_f]{\bf \Psi_i}[\chi_i]\prod_t {\cal D}(\f^{(t)}) {\cal D}(g_{\m\n}^{(t)}) e^{i[{\cal S}_\f (\f, g_{\m\n},\F_j^c)+ {\cal S}_g (g_{\m\n},\F_j^c]}\, ,
\end{equation}
where ${\cal S}_\f =0$ for $\f =0$.  Consider metric fluctuations around some background $g_{\m\n}^0$ that will be later determined.  Writing $g_{\m\n}= g_{\m\n}^0 + h_{\m\n}$, we expand the action in Eq.(\ref{field}) in powers of $h_{\m\n}$
\begin{equation}
{\cal S}={\cal S}_\f^0 (\f, g_{\m\n}^0,\F_j^c)+ {\cal S}_g^0 (g_{\m\n}^0,\F_j^c)+  \int\left(\frac{\d {\cal S}_g }{\d g_{\m\n}}\right)_0 h_{\m\n}+ \int\left(\frac{\d{\cal S}_\f }{\d g_{\m\n}}\right)_0 h_{\m\n} + \cdots
\end{equation}
The  path integral for the Green functional Eq.(\ref{Greenfunct}) reads
\begin{eqnarray}
\label{1hmn}
U[\f_i,\f_f;g_{\m\n}^0] =e^{i{\cal S}_g^0}&&\left(\int_{\phi^{(i)}}^{\phi^{(f)}} \prod_t{\cal D}(\f) e^{i{\cal S}_\f^0}\right) \int\prod_t {\cal D}(h_{\m\n})\exp \left[\int i \left(\frac{\d {\cal S}_g }{\d g_{\m\n}}\right)_0 h_{\m\n} + \cdots \right]\nn \\ &&\left [1+\frac{i\int_{\phi^{(i)}}^{\phi^{(f)}} \prod_t {\cal D}(\f) e^{i{\cal S}_\f^0}\int\left(\frac{\d {\cal S}_\phi }{\d g_{\m\n}}\right)_0}{\int_{\phi^{(i)}}^{\phi^{(f)}} \prod_t{\cal D}(\f)e^{i{\cal S}_\f^0}}\, h_{\m\n}+ \cdots\right]\, .
\end{eqnarray}
We  changed the integration variables from $g _{\mu\nu}$  to $h_{\mu\nu}$ and integrated between the configurations $h^{(i)}_{\mu\nu}=h^{(f)}_{\mu\nu}=0$. 

We tentatively assume that the gravitational fluctuations are small and perform a first order analysis, whose validity will be discussed later. Accordingly, we take  functionals ${\bf \Psi _i} $ and ${\bf \Psi _f} $ in Eq.(\ref{field}) peeked on the configurations  $h^{(i)}_{\mu\nu}=h^{(f)}_{\mu\nu}=0$ with negligible spread, so that Eq.(\ref{field}) can be written as
\begin{equation}
\label{field1}
\langle f \vert i \rangle = \int{\cal D}\f_f{\cal D}\f_i  {\boldsymbol \psi _f}^*[\f_f]{\boldsymbol \psi_i}[\f_i]U[\f_i,\f_f;g_{\m\n}^0] \, ,
\end{equation}
where  
\begin{equation}
\label{defpsi}
\boldsymbol{ \psi }_i[\phi_i]=\int {\cal D}(g^{(i)}_{\mu\nu})\boldsymbol{ \Psi }_i[\chi_i]\quad,\quad \boldsymbol{ \psi }_f[\phi_f]=\int { {\cal D}}(g^{(f)}_{\mu\nu})\boldsymbol{ \Psi }_f[\chi_f]\, .
\end{equation}
In absence of gravitational degrees of freedom we get
\begin{eqnarray}
\label{i}
\langle f \vert i \rangle_0&=&\int{\cal D}(\phi^{(i)}){\cal D}(\phi^{(f)}) {\boldsymbol{ \psi }_f}^*[\phi_f]{\boldsymbol{ \psi }_i}[\phi_i]\prod_t{\cal D}(\f) e^{i{\cal S}_\f^0}\\
&:=&\int \boldsymbol{ {\cal D}}(\f){\boldsymbol{ \psi }_f}^*[\phi_f]{\boldsymbol{ \psi }_i}[\phi_i]  e^{i{\cal S}_\f^0}\, .\label{shortnote}
\end{eqnarray}
We introduced in Eq.(\ref{shortnote}) the condensed writing $\boldsymbol{ {\cal D}}$ for an integration on the all slicing of the space, including the boundary slices. The functional integral Eq.(\ref{field}) becomes
\begin{eqnarray}
\label{1}
\langle f \vert i \rangle=e^{i{\cal S}_g^0}\  \langle f\vert i\rangle_0&&\int \boldsymbol{ {\cal D}}(h_{\m\n})\exp i\int \left[ \left(\frac{\d {\cal S}_g }{\d g^{\m\n}}\right)_0\right . \nn\\
 &&+\left .\frac{\int\boldsymbol{ {\cal D}}(\phi) {\boldsymbol{ \psi }_f}^*[\phi_f]{\boldsymbol{ \psi }_i}[\phi_i]   e^{i{\cal S}_\f^0}\int\left(\frac{\d {\cal S}_\phi }{\d g_{\m\n}}\right)_0 }{\int \boldsymbol{ {\cal D}}(\f){\boldsymbol{ \psi }_f}^*[\phi_f]{\boldsymbol{ \psi }_i}[\phi_i]  e^{i{\cal S}_\f^0}}\right] h_{\m\n}+ \cdots 
\end{eqnarray}
where the omitted terms are second order or higher in $h_{\m\n}$.

Up to now the choice of the background  was left unspecified. We define a background by a saddle point  in  Eq.(\ref{1}), whether or not the steepest descent evaluation constitutes a good approximation of the path integral.  Such unrestricted backgrounds may depart considerably from the usual classical limit. The vanishing of first order terms yields
\begin{equation}
\label{zero}
\langle f\vert i\rangle= C\, e^{i{\cal S}_g^0}\, \langle f\vert i\rangle_0 + \cdots\,
\end{equation}
where C is a determinant and $\langle f\vert i\rangle_0$  the amplitude of the quantum matter evaluated in the background $g_{\m\n}^0$. The dots refer to cubic or higher order terms in $h_{\m\n}$. 
Taking ${\cal S}_g$ to be the Einstein action we get
\begin{equation}
\label{steep}
E_{\m\n}^0\equiv R_{\m\n}^0-\frac{1}{2} g_{\m\n}^0 R^0 - 8\pi \,T_{\m\n}^0(\F_j^c) = 8\pi  \frac{\langle f\vert \hat T_{\m\n}^\f\vert i\rangle_0}{\langle f\vert i\rangle_0}\, .
\end{equation}
Here $T_{\m\n}^0(\F_j^c)$ is the energy-momentum tensor of the classical matter in the background $g_{\m\n}^0$ and $ \hat T_{\mu\nu}^\f$ is, in the Heisenberg representation, the energy-momentum tensor operator of the quantum matter $\f$ in this background. Explicitly, 
 \begin{equation}
 \label{explicit}
 \frac{\langle f\vert \hat T_{\m\n}^\f\vert i\rangle_0}{\langle f\vert i\rangle_0}=\frac{2}{\sqrt{- g}}\frac{\int\boldsymbol{ {\cal D}}(\phi) {\boldsymbol{ \psi }_f}^*[\phi_f]{\boldsymbol{ \psi }_i}[\phi_i]   e^{i{\cal S}_\f^0}\left(\frac{\d {\cal S}_\phi }{\d g^{\m\n}}\right)_0 }{\int  \boldsymbol{ {\cal D}}(\f) {\boldsymbol{ \psi }_f}^*[\phi_f]{\boldsymbol{ \psi }_i}[\phi_i] e^{i{\cal S}_\f^0}}\, .
 \end{equation}
 Eqs.(\ref{steep}) and (\ref{explicit}) show that the detailed back-reaction of the quantum matter  on the gravitational background is driven by a functional of $\f$, labelled the weak value of  $ \hat T_{\mu\nu}$~\cite{AC}.   Eq.(\ref{steep}) should be viewed as an equation for a (generally complex) background.  If  higher order quantum corrections, that is higher order terms in Eq.(\ref{zero}), may be disregarded, we shall qualify the  background  as semi-classical. In that case, 
the boundary configurations in the amplitude Eq.(\ref{1}) must be taken to be consistent with the boundary conditions imposed on  the solution of the integro-differential equation Eq.(\ref {steep}).

 It is quite remarkable that the quantum nature of the matter field on a background is encoded in the weak value of the energy-momentum tensor, which, as will be shown in Section~3, may depart in an important way from its expectation value. This has in particular the following consequence. Consider a classical background metric $g_{\m\n}^{\bar 0}$ solving the classical Einstein equations:
\begin{equation}
\label{clasbg}
 R_{\m\n}^{\bar 0}-\frac{1}{2}  g_{\m\n}^{\bar 0} R^{\bar 0}- 8\pi \,  T_{\m\n}^{\bar 0}( \F_j^{\bar c})=0 \, .
\end{equation}
and define $\delta g^{f i}_{\m\n}$ as the first order correction induced by the last term of Eq.(\ref{steep}). Expanding the left hand side of Eq.(\ref{steep}) around $  g_{\m\n}^0=  g_{\m\n}^{\bar 0} + \delta g^{f i}_{\m\n}$, we obtain a linear equation:
\begin{equation}
\label{pertgmn}
{\cal A}_{\mu\nu}^{\rho\sigma}\,\delta g^{f i}_{\rho\sigma} = 8\pi  \frac{\langle f\vert \hat T_{\rho\sigma}^\f\vert i\rangle_{\bar 0}}{\langle f\vert i\rangle_{\bar 0}}\, .
\end{equation}
Here,   we are looking for a solution of Eqs (\ref{clasbg}) and (\ref{pertgmn}) interpolating between the geometries given on $\S_0$ and $\S_1$. Let us emphasize that we assume its existence and unicity as part of the definition of  the background $  g_{\m\n}^0$. We may then interpret these solutions as the matrix elements of a metric perturbation operator
\begin{equation}
\label{soldgmn}
\delta g^{f i}_{\rho\sigma} =   \frac{\langle f\vert   \delta \hat g_{\rho\sigma} \vert i\rangle_{\bar 0}}{\langle f\vert i\rangle_{\bar 0}}\, ,
\end{equation}
which, accordingly, is completely defined with respect to the background metric $g_{\m\n}^{\bar 0}$. The linearized gravity response to quantum matter encoded in Eqs.(\ref{pertgmn}) and (\ref{soldgmn}) will play an essential r\^ole in the  analysis of Section~3.

We now review the properties of weak values which are relevant in the black hole context.

\subsection{Exclusive and inclusive backgrounds}
The evolution of a  quantum system is determined once the initial or ``pre-selected'' Schr\"odinger state $\vert i_{t_i}\rangle$ at time $t_i$ is fixed. More information can be obtained on the system if one fixes in addition a final or ``post-selected'' state $\vert f_{t_f}\rangle$ at time $t_f$. Such time-symmetric presentation of quantum physics was introduced by Aharonov et al.~\cite{ABL: 1964}.  The weak value  of an operator $\hat A$ at some intermediate time $t$ was defined in reference ~\cite{AC}. It reads, in the Heisenberg representation,
\begin{equation}  
\label{weak}
A_{[weak]}^{f i}(t) = \frac{\langle f \vert \hat
A(t) \vert i \rangle}{\langle f \vert i\rangle}\, ,
\end{equation}
where we labelled $\vert i\rangle $ the in-state and $\vert f\rangle$ the out-state. 
In Section 4, we shall consider the mass $M$ black hole S-matrix between an in-state defined by the functional of configurations $\vert i\rangle $ on $\cal J^-$ describing its constituents and  out-states $\vert f\rangle$ describing its decay products on $\cal J^+$, in analogy with an exclusive S-matrix element $\langle f\vert i\rangle$ in elementary particle scattering processes. 
 More generally, in what follows, we shall consider for a given in-state  $\vert i\rangle $ a complete set of orthonormal  decay products out-states $\vert f\rangle $ such that $0<\vert\langle i \vert f \rangle\vert <1$, and refer to such amplitudes as to exclusive ones.  Using the completeness relation and Eq.(\ref{weak}), we see that the weak values are related to the expectation value of  $\hat A$ by
\begin{equation}
\label{weakav}
\sum_f p_f\,A_{[weak]}^{f i}=\langle i \vert \hat A \vert i \rangle \quad,\quad   p_f =\vert\langle f \vert i\rangle \vert^2 \, .
\end{equation}

 We now consider more generally the case where  the out-state $\vert f\rangle$ is left partially or totally unspecified. To this effect, we first introduce generalized weak values~\cite{MP, Brout:1995rd}. 

The set of out-states span some Hilbert space $\mathbf H$. Let us consider a decomposition of $\mathbf H$ into a tensor product ${\mathbf H}= {\mathbf H}_1\otimes {\mathbf H}_2$. We perform a ``partial'' post-selection on ${\mathbf H}_1$, $\vert f_{{\mathbf H}_1}^j\rangle \in {\mathbf H}_1$, leaving the final state on ${\mathbf H}_2$ unspecified. Writing $A_{[weak]}^{fi}  $ in  Eq.(\ref{weak}) as 
\begin{equation}
\label{partial0}
A_{[weak]}^{fi}  = \frac{tr \vert f \rangle \langle f \vert \hat A \vert i \rangle \langle i \vert}{ tr \vert f \rangle \langle f \vert i\rangle  \langle i \vert}\, ,
\end{equation}
 one generalizes the weak value to
\begin{equation}
\label{partial1}
A_{[weak]}^{f_{{\mathbf H}_1}^j\, i}  = \frac{tr {\Pi_{{\mathbf H}_2}}\vert f_{{\mathbf H}_1}^j\rangle \langle f_{{\mathbf H}_1}^j\vert \hat
A \,\vert i \rangle \langle i \vert}{ tr {\Pi_{{\mathbf H}_2}} \vert f_{{\mathbf H}_1}^j\rangle \langle f_{{\mathbf H}_1}^j\vert i\rangle  \langle i \vert}\, ,
\end{equation}
where $ {\Pi_{{\mathbf H}_2}}$ is the projection operator onto ${\mathbf H}_2$. Equivalently one has
\begin{equation}
\label{partial}
A_{[weak]}^{f_{{\mathbf H}_1}^j\, i}   = \frac{\langle i\vert f_{{\mathbf H}_1}^j\rangle \langle f_{{\mathbf H}_1}^j\vert \hat
A \,\vert i \rangle}{{\langle i} \vert f_{{\mathbf H}_1}^j\rangle \langle f_{{\mathbf H}_1}^j\vert i\rangle}\, .
\end{equation}
Partial post-selection on ${\mathbf H}_1$ is thus equivalent to the choice of a final state $\vert f_{{\mathbf H}_1}^j \rangle \langle f_{{\mathbf H}_1}^j\vert i\rangle$, that is of a EPR conjugate pair $\vert f_{{\mathbf H}_1}^j\rangle$ on ${\mathbf H}_1$ and $\langle f_{{\mathbf H}_1}^j\vert i\rangle$ on ${\mathbf H}_2$. Using the completeness relation in Eq.(\ref{partial}), we see that the generalized weak values are related to the expectation value of  $\hat A$ by\footnote{The states $\vert i\rangle,\vert f\rangle$ and  $\vert f_{{\mathbf H}_1}^j\rangle$ are normalized to one. Generically, the EPR conjugate state of the latter, written as $\langle f_{{\mathbf H}_1}^j\vert i\rangle$, is not.} 
\begin{equation}
\label{partialav}
\sum_j p_{f_{{\mathbf H}_1}^j}\,A_{[weak]}^{f_{{\mathbf H}_1}^j\, i}   =\langle i \vert \hat A \vert i \rangle \quad, \quad  p_{f_{{\mathbf H}_1}^j}= {\langle i} \vert f_{{\mathbf H}_1}^j\rangle \langle f_{{\mathbf H}_1}^j\vert i\rangle \, .
\end{equation}
When the final state is left completely unspecified, namely  if no post-selection is performed,  the projection $ {\Pi_{{\mathbf H}_2}}$ in Eq.(\ref{partial1}) extends to the full Hilbert space ${\mathbf H}$ and one obtains
\begin{equation}
\label{nopost}
A_{[weak]}^{ \, i}  = \frac{tr \Pi_{\mathbf H}\hat
A \,\vert i \rangle \langle i \vert}{ tr \Pi_{\mathbf H} \vert i\rangle  \langle i \vert} =\langle i\vert  \hat
A(t) \,\vert i \rangle\, ,
\end{equation}
recovering the expectation value as the generalized weak value of $ \hat A$ in absence of post-selection.

The real part of the weak value  of an observable  $\hat A$, Eqs (\ref{weak}), was interpreted in reference~\cite{AC} as the result of  its weak, {\it i.e.} approximate, measurement at times intermediate  between the pre- and post-selections by  exact measurements of the states  $\vert i\rangle$ and  $\vert f\rangle$.  In the present work, weak values of the energy-momentum tensor $\hat T_{\m\n}$ appear in Eq.(\ref{steep}) as sources of quantum matter back-reaction on background geometries. Such backgrounds will be labelled ``exclusive backgrounds''. The weak values of the energy-momentum tensor operator 
act in Eq.(\ref{steep}) as quantum sources for background geometries in exclusive  amplitudes, namely for exclusive backgrounds. 

The fully inclusive amplitude $\langle i\vert i\rangle$, $\langle i_{t_f}\vert U(t_f, t_i)\vert i_{t_i}\rangle$ in the Schr\"odinger representation, is just unity.  From the functional representative of $\vert i_{t_f}\rangle$ in the path integral, on may select a single saddle-point 
that yields from Eqs (\ref{steep}) and (\ref{nopost}) a real metric background driven by the expectation value  $T_{\mu\nu[weak]}^{\,i}   =\langle i \vert \hat T_{\mu\nu} \vert i \rangle $. We label it the ``inclusive background''.
 Alternatively unitarity allows to write the inclusive amplitude  as a coherent superposition of exclusive amplitudes, namely $\sum_f a_{fi} \langle f\vert i\rangle$ with  $a_{fi}=  \langle i\vert f\rangle$. As a consequence of Eq.(\ref{weakav}), the inclusive background appears as  a statistical average of the exclusive ones
\begin{equation}
\label{average}
E_{\m\n}^0(g_{\m\n})=\sum_f \vert \langle i  \vert f\rangle\vert^2 \, E_{\m\n}^{0}(g_{\m\n}^{f i})\, ,
\end{equation}
where the sum is  over the (possibly complex) backgrounds $g_{\m\n}^{f i}$ for  a complete set of post-selected states $\vert f\rangle$ and a suitable initial state.  In the linear approximation Eq.(\ref{soldgmn}) around some average background it reduces to $\d g_{\m\n}= \sum_f   \vert \langle i  \vert f\rangle\vert^2\,  \d g_{\m\n}^{fi}$.  This averaging  results from the denominator in the weak values driving both inclusive and exclusive backgrounds. Let us emphasize that, as will be apparent later,  on $\S_1$, the representatives of $\vert f\rangle$  may differ wildly from that of  $\vert i\rangle$. 

Partial post-selections define from Eq.(\ref{steep}), using in its right hand side the corresponding generalized weak values, ``semi-inclusive'' backgrounds. We summarize in Table 1 the relation between the quantum sources of exclusive and inclusive backgrounds. 

The generalized weak values of the energy-momentum tensor operator 
act in Eq.(\ref{steep}) as quantum sources for both exclusive and (semi-)inclusive backgrounds.  We shall discuss in the Section 3, for times short compared to black hole lifetime, inclusive and semi-inclusive backgrounds resulting from post-selection on $\cal J^+$, taking as unperturbed metric the one  depicted by the Penrose diagram of Fig.1. We shall  show that post-selection induces huge metric fluctuations in the vicinity of the horizon.  In Section 4, we  shall argue that similar effects are present in the S-matrix when back-reaction is fully taken into account and that they provides a clue for solving the black hole information paradox.

\begin{table}[h]
\caption{{\sl \small Links between the quantum sources of metric backgrounds for exclusive and inclusive backgrounds. In the linear approximation around a given background metric $g_{\m\n}^{\bar 0}$ these relations imply similar ones between metrics corrections $ \delta g^{f i}_{\r\s}$ in Eq.(\ref{soldgmn}).}}
\begin{center}
$\left|
\begin{array}{c|c|c}
\hline
\hbox {post selection} &\sum_f p_f\,T_{\mu\nu[weak]}^{f i}=\langle i \vert \hat T_{\mu\nu} \vert i \rangle    & \hbox {exclusive} \\ \hline 
\hbox {partial post selection} &\sum_j p_{f_{{\mathbf H}_1}^j}  T_{\mu\nu[weak]}^{f_{{\mathbf H}_1}^j\, i}   =\langle i \vert \hat T_{\mu\nu} \vert i \rangle    & \downarrow  \\
  \hline
\hbox {no post selection} &T_{\mu\nu[weak]}^{\,i}   =\langle i \vert \hat T_{\mu\nu} \vert i \rangle    &  \hbox {inclusive} \\ \hline
\end{array}
\right| $
\end{center}
\end{table}

\section{Back-reaction of the Hawking radiation}

\subsection{The black hole paradox : singularity and horizon}
The ``conventional" description of black hole evaporation, depicted schematically in Fig.2, leaves us with the alternative of violating unitarity or rejecting as fundamentally incorrect the Hawking derivation of black hole radiation. This alternative is rooted in the assumption that the unitarity issue originates in the problem posed by the singularity. However unitarity may possibly be related to the fate of the  horizon:   the conventional derivation of the radiation requires transplanckian quantum fluctuations close to the horizon and the question arises whether these induce important back-reaction.

\begin{figure}[h]
   \centering
   \includegraphics[width=16 cm]{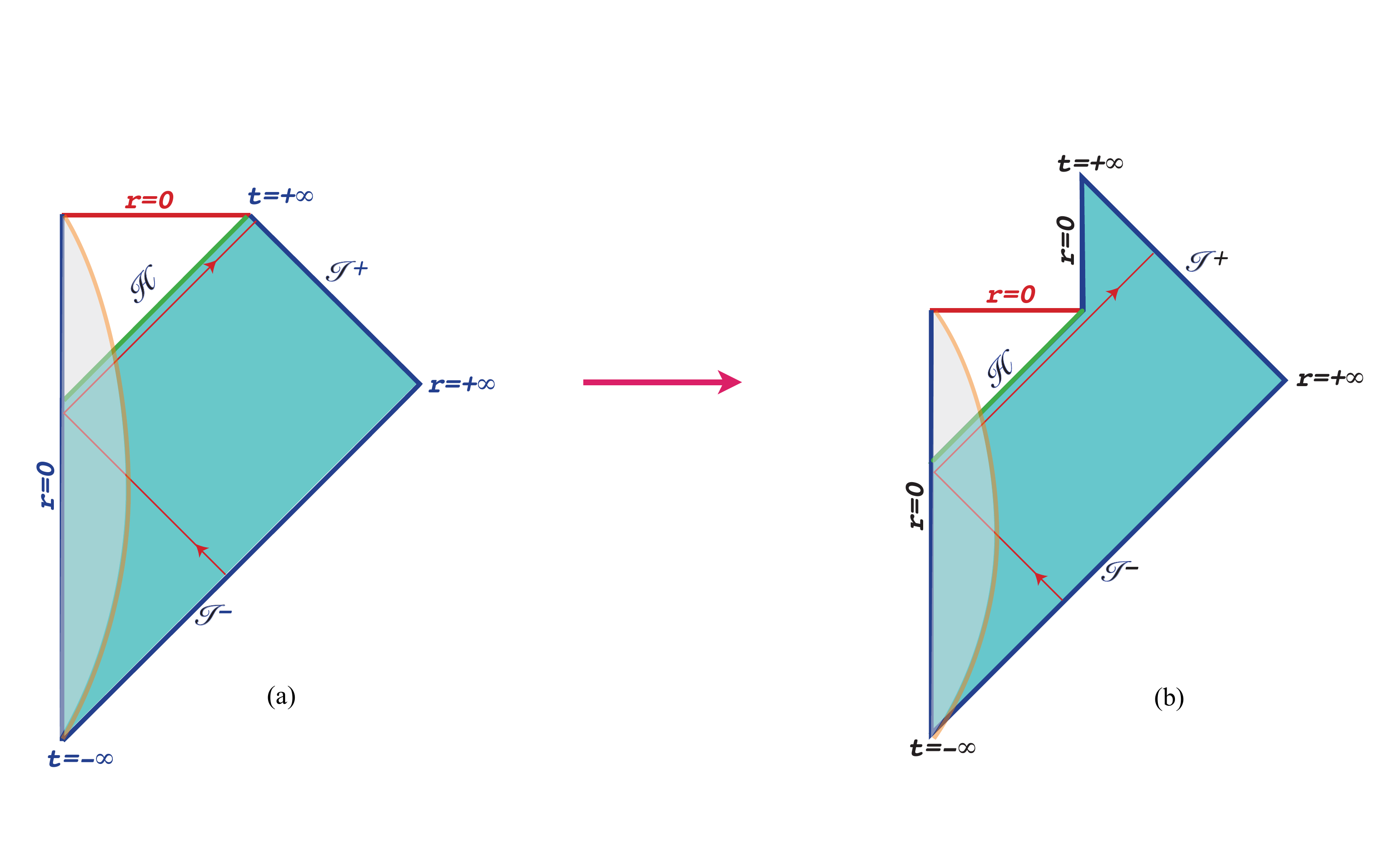} 
 \caption { \sl \small ``Conventional" Penrose diagrams of a collapsing shell. (a) Without back reaction : the horizon grows from 0 to 2M at the star surface and remains constant outside the star. The trajectory of an s-wave Hawking light ray is depicted in red. The region outside the horizon is painted in blue and the Minkowskian region inside the star in gray. (b) Including back-reaction : the horizon grows from 0 to   2M at the star surface, then decreases to 0 during the evaporation.} 
\end{figure}

At first sight, the answer is negative. The average energy momentum tensor in the background of Fig.1  is completely insensitive to transplanckian fluctuations, and the equivalence principle strongly suggests that the back-reaction felt by a  classical free-falling object does not affect significantly its trajectory across the horizon. Moreover a  quantum particle crossing the horizon does not experience a dramatic gravitational interaction from these fluctuations because the contributions of the transplanckian ancestors of the Hawking photons are cancelled by those of their partners on the other side of the horizon~\cite{giddings}. Again, this is not unexpected, as the small curvature near the horizon makes the black hole event horizon similar to the Rindler horizon which should not prevent a particle to travel freely across it in Minkowski space. Thus,  for times short compared to the black hole lifetime, the conventional background of Fig.1 appears as a valid classical background in which quantum effects, including gravitational back-reaction, take place.

Our point is that those arguments are not sufficient to ensure that the horizon survives the Hawking radiation back-reaction. In absence of post-selection the classical geometry of Fig.1 can be used as a good approximation for the inclusive background. We shall confirm in Section 3.2 that  the back-reaction  remains weak in the vicinity of the horizon. But we shall show in Section~3.3 that a generic post-selection of a state on $\cal J^+$ induces a strong back-reaction, questioning the very existence of the classical horizon for semi-inclusive backgrounds. In this perspective, the black hole unitarity issue may be rooted in the horizon rather than in the classical singularity, as a horizon-free black-hole S-matrix could be consistent with unitarity~\cite{'tHooft:1984re,Stephens:1993an}. 

We  determine here the first order effect of quantum matter on the classical geometry of the incipient black hole depicted in Fig.1. The impact of post-selection on the quantum matter energy-momentum tensor is reviewed~\cite{MP, Englert:1994qe} and further analyzed. We derive in the linear approximation the corresponding back-reaction on the background geometry and discuss its consequences. We shall then argue in Section 4 that the essential features uncovered in this simple case have a more general validity.   

\subsection {Back-reaction of the Hawking radiation: no post-selection}
The background of Fig.1  depicts the classical space-time in which the Hawking radiation is computed in absence of back-reaction. If no post-selection is performed, the back-reaction for the inclusive background is driven by the average energy-momentum tensor. Performing a suitable renormalization, the latter is regular on the horizon in the free-falling frame as is characteristic of the ``Unruh vacuum''. To prepare for the discussion of the effect of post-selection, we first review the computation the average energy-momentum tensor,  stressing the link between horizon regularity and Hawking radiation.

Consider first the s-wave contribution of the field $\f$, neglecting the small residual relativistic barrier (see Eq.(\ref{radial4}) below). This amounts to truncate the metric  Eq.(\ref{Swartz}) to the two-dimensional one
\begin{equation}
ds^2= \left(1- \frac{2M}{r }\right) du\  dv \label{Swarz2}\, ,
\end{equation}
together with Eqs.(\ref{uv}) and (\ref{tortoise}). The Heisenberg equation of motion for $\f$ is 
\begin{equation}
\label{phi}
\partial_u\partial_v \f =0 \, ,
\end{equation}
and the trace anomaly yields
\begin{equation}
\label{trace}
\langle i\vert \hat T_{uv}^{(2)}\vert i\rangle =-\frac{1}{96\pi}(1- \frac{2M}{ r }) R^{(2)} =-\frac{1}{24\pi}\frac{M}{r^3}(1- \frac{2M}{ r } ) \, ,
\end{equation}
where $R^{(2)}$ is the two-dimensional scalar curvature and the expectation value of $ \hat T_{uv}^{(2)}$ is related to the (s-wave) contribution to the expectation value of the four-dimensional energy-momentum tensor\footnote{We distinguish  two- and four-dimensional energy-momentum  and curvature tensors by a superscript $(2)$ for the former and none for the latter.} $\hat T_{uv}$ by
$\hat T_{uv}^{(2)}= 4\pi r^2  \,\hat T_{uv}$.
From Eq.(\ref{trace}), integrating the conservation laws for $ \hat T_{uu}^{(2)}$ and $ \hat T_{vv}^{(2)}$, one gets outside the star
\begin{eqnarray}
\label{Tvv}
&&\langle i\vert \hat T_{vv}^{(2)} \vert i\rangle ={1\over 12\pi} \left[-{M\over
2r^3}(1-{2M\over r}) - {M^2\over 4r^4}  \right]  + t_v(v) \, ,  \\ 
\label{Tuu}
&&\langle i\vert \hat T_{uu}^{(2)} \vert i\rangle ={1\over 12\pi} \left[-{M\over
2r^3}(1-{2M\over r}) - {M^2\over 4r^4}  \right] + t_u(u)\, .   
\end{eqnarray}
One ensures Minkowskian boundary conditions on ${\cal J}^-$ by requiring $t_v(v)=0$. This is equivalent in field theory to a renormalization of the energy-momentum tensor by subtracting its local (divergent) Minkowskian expectation value. Consistency with Hawking radiation then imposes 
\begin{equation}
\label{tu}
\lim_{u\to +\infty} t_u(u)= \frac{1}{2\pi}\int_0^\infty \frac {\o}{e^{8\pi M\o}-1}d\o =\frac{1}{(4M)^2}\frac{1}{48\pi} \, .
\end{equation}
 One gets in the vicinity of the horizon for asymptotically large advanced times ($r\to 2M$, $v\rightarrow +\infty$)\footnote{For finite $v$, the first non-vanishing term  in the expansion of $ t_u(u)$ in terms of $(1-2M/r)$ is of order $(1-2M/r)^2$ and has the form $A(1/M^2)(1-2M/r)^2\exp(-v/4M)$ where A is a numerical constant (for a collapsing light-like shell along $v=v_s$, $A=-\exp (v_s/4M)/128 \pi M^2$). This terms yields thus for finite $v$ a contribution to $\langle i\vert \hat T_{uu}^{(2)}\vert i\rangle$ of the same order in $(1-2M/r)$ as the leading contribution of $ t_u(u)$ but vanishes on $\cal J^+$. In what follows, when expressing matrix elements of $\hat T_{uu}$ in the vicinity of the horizon, we shall {\em always} quote the value obtained from the leading order of $t_u(u)$, hence its expression on $\cal J^+$. As in Eq.(\ref{TuuH})  hereafter, this ensures for asymptotic large $v$ the validity of the expression up to $r\lesssim O(M)$. Quantitative results for $v$ finite differ only by a finite numerical factor which does not affect its convergent (or divergent) character on the horizon.},
\begin{eqnarray}
\label{TvvH}
\langle i\vert \hat T_{vv}^{(2)}\vert i\rangle &=& {1\over 12\pi} \left[-{M\over
2r^3}(1-{2M\over r}) - {M^2\over 4r^4}  \right] \stackrel{r\to 2M}{\longrightarrow}\frac{1}{(4M)^2}\frac{-1}{48\pi}\, ,\\\nn\\ 
\label{TuvH}
\langle i\vert \hat T_{uv}^{(2)}\vert i\rangle &=&-\frac{1}{24\pi}\frac{M}{r^3}(1- {2M\over r }) \stackrel{r\to 2M}{\longrightarrow}(1- {2M\over r }) \frac{1}{(4M)^2}\frac{-1}{12\pi}\, ,\\
\nn\\
\nn
\langle i\vert \hat T_{uu}^{(2)}\vert i\rangle &=& {1\over 12\pi} \left[-{M\over
2r^3}(1-{2M\over r}) - {M^2\over 4r^4}  \right] + \frac{1}{(4M)^2}\frac{1}{48\pi}\\
 \label{TuuH}&&\,\qquad\qquad\qquad\qquad\stackrel{r\to 2M}{\longrightarrow} \frac{1}{(4M)^2}\frac{1}{12\pi}(1-{2M\over r})^2\, .
\end{eqnarray}
Consider the energy-momentum tensor felt by an object heading towards the horizon on a geodesic trajectory. On such trajectory
\begin{eqnarray}
\label{geou}
\frac {du}{ds} &\stackrel{r\to 2M}{\longrightarrow}&[C (1-{2M\over r})]^{-1}\, ,\\
\label{geov}\frac {dv}{ds} &\stackrel{r\to 2M}{\longrightarrow}&C \, ,
\end{eqnarray}
 where $C$ is an integration constant. From Eqs.(\ref{TvvH}),(\ref{TuvH}),(\ref{TuuH}), and (\ref{geou}),(\ref{geov}), we see that  $T_{\m\n}^{(2)}u^\m u^\n \equiv T_{\t\t}^{(2)}$, where $\t$ is the proper time of the object, is regular on the horizon outside the star, vanishes in the limit $M\to\infty$, and is not positive definite. One gets
\begin{equation}
\label{unruh}  
T_{\t\t}^{(2)}= \frac{1}{(4M)^2}\frac{1}{12\pi}\left(C^{-1}-\frac{2+\sqrt{5}}2C\right)\left(C^{-1}-\frac{2-\sqrt{5}}2C\right)\, .
\end{equation}
The regularity of the energy in the free falling frame Eq.(\ref{unruh}) is a property of the Unruh vacuum.
Alternatively, one may characterize the behavior of the average energy-momentum tensor outside the star in the vicinity of the horizon in a coordinate invariant way. One computes the  renormalized scalar $ \langle T^2\rangle ^{(2)}_{Unruh} $ defined by
\begin{equation}
\label{T2}
\langle T^2\rangle ^{(2)}_{Unruh} =T_{\m\n}^{(2)}T^{\m\n\,{(2)}}=2T_{uu}^{(2)}(g^{uv})^2T_{vv}^{(2)}+2T_{uv}^{(2)}(g^{uv})^2T_{uv}^{(2)}\, ,
\end{equation}
 where $T_{\m\n}^{(2)} \equiv \langle i \vert \hat T_{\m\n}^{(2)}\vert  i\rangle$.
Using Eqs.(\ref{TvvH}), (\ref{TuvH}) and (\ref{TuuH}), we recover on the horizon a finite contribution that vanishes in the $M\to\infty$ limit:
\begin{equation}
\label{2hor}
\langle T^2\rangle ^{(2)}_{Unruh} \stackrel{r\to 2M}{\longrightarrow}\frac{1}{(4M)^4}\frac{1}{24\pi^2}\, .
\end{equation}

In the four dimensional theory, $T_{\m\n}$ in the free falling frame is also finite on the horizon and therefore so is the scalar $\langle T^2\rangle_{Unruh}$. Its s-wave contribution is, except for the small relativistic barrier, given by  $\langle T^2\rangle ^{(2)}_{Unruh} /\pi^2 M^4 $. Such a finite contribution persists when all modes are taken into account~\cite{Candelas}
\begin{equation}
\label{4hor}
\langle T^2\rangle_{Unruh} \stackrel{r\to 2M}{\longrightarrow}O\left(\frac{1}{M^8}\right) .
\end{equation}
To characterize the linear back-reaction Eq.(\ref{pertgmn}) in absence of post-selection, we consider Eq.(\ref{steep}) with the weak value on the r.h.s. taken to be the expectation value evaluated  in the Schwarzschild background $g_{\m\n}^{\bar 0}$ given in Eq.(\ref{Swartz}). Outside the star we put $T_{\m\n}^0(\{ \F_j^{ c}\})=0 $ to get 
\begin{equation}
\label{backunruh}
R_{\m\n}^0 R^{\m\n \,0} = 64 \pi^2 \langle T^2\rangle_{Unruh}\, .
\end{equation}
Thus if no-post selection is performed, the back-reaction to the  inclusive background remains small in the vicinity of the unperturbed background and the curvature at its horizon remains small and vanishes when $M\to \infty$. The Penrose diagram of Fig.1 constitutes  a valid approximation to the inclusive  background for times  small compared to the black hole lifetime. 

\subsection {Back-reaction of the Hawking radiation: partial post-selection}
\subsubsection{A simple model}
 We have depicted in Fig.1 the light ray trajectory of an s-wave ending on ${\cal J}^+$ and that of its partner ending in the singularity on the other side of the horizon $\cal H$. However that part of space-time is irrelevant for the derivation of the Hawking radiation from an incipient black hole. It can indeed be derived solely from the region (painted in blue in Fig.1)  outside of the event horizon. The Hawking radiation is sometimes interpreted, not only as a pair creation but also as the tunneling of the partner of the Hawking quantum through the horizon. This interpretation can be made quantitative~\cite{PW} but involves already some back-reaction, as  energy-conservation, including the black hole mass, must be imposed. In the present work, we consider the dependence of the back-reaction on the out-states and we wish to keep only the minimal hypothesis necessary to derive the Hawking radiation. This  does not require the inclusion of the space-time beyond the horizon. For sake of completeness, we briefly recall here the derivation for the shell model which contains all the necessary ingredients.

\vskip 1cm
 \begin{figure}[h]
   \centering
   \includegraphics[width=5 cm]{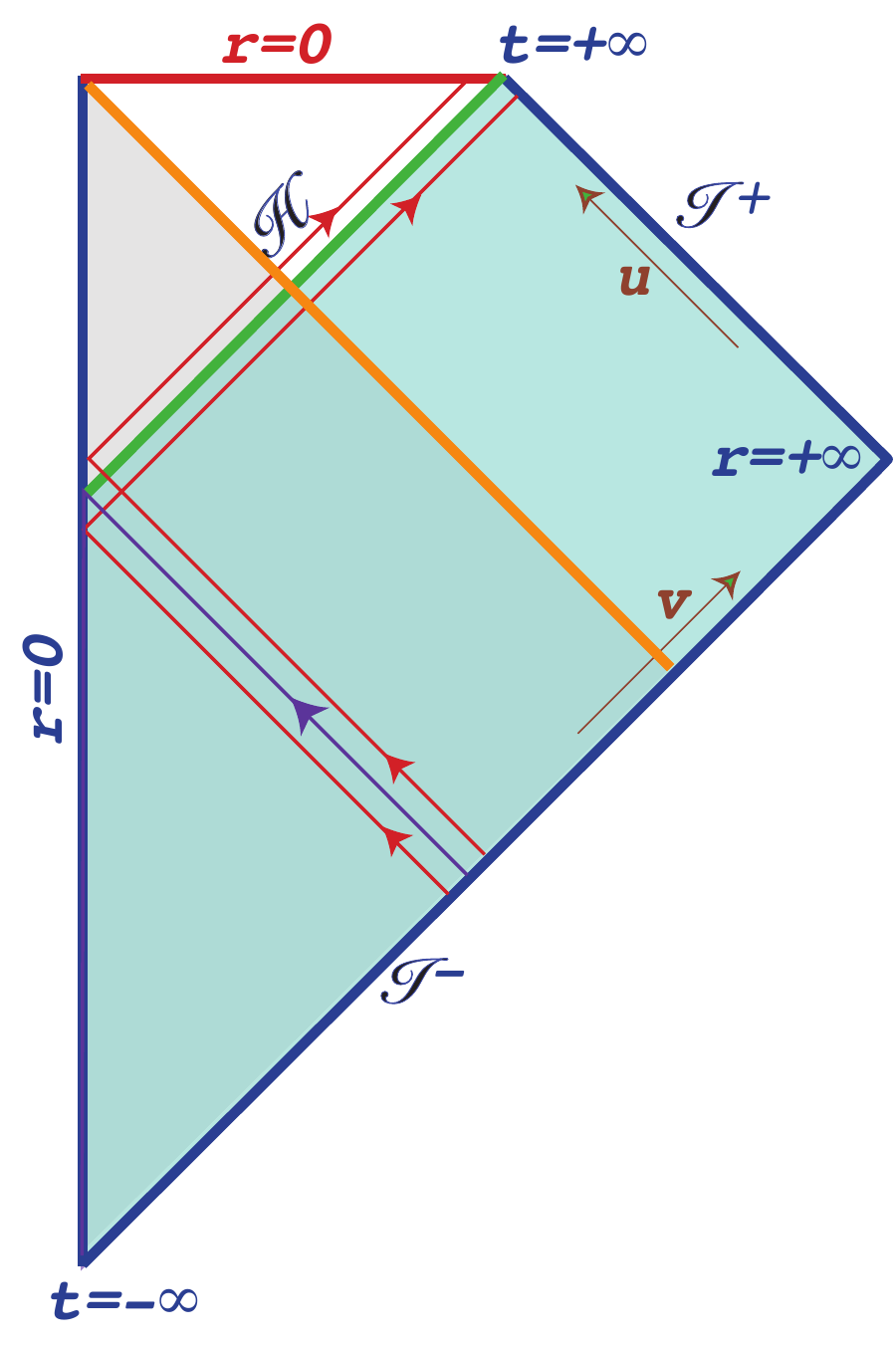} 
 \caption { \sl \small The Penrose diagram of Fig.1 for a light-like shell.} 
\end{figure}
\vskip 1cm

As in Section 3.2, we first limit ourselves to s-waves and neglect the residual relativistic barrier. In addition we shall take a light-like shell in order to avoid irrelevant constants in the near-horizon wave-functions. The Penrose diagram is depicted in Fig.3. When $r_s-2M\le O(M)$ where $r_s$ is the star radius, one may use in Eq.(\ref{Swarz2}) the approximation 
 \begin{equation}
\left(1- \frac{2M}{r_s }\right)\simeq \exp\frac{(v_s-u)}{4M}
\label{approx}\, ,
\end{equation}
where $v_s$ is the advanced time on the shell. In this limit,  we take a solution of the field equation Eq.(\ref{phi}) expressed in terms of a complete set of normed wave functions $\vert \pm \o)$ on ${\cal H}\cup{\cal J}^+$. These are
\begin{eqnarray}
\Phi(u,v)&=&\int_0^\infty  d\omega \left[\vert
-\omega^{out}) a^{out}_{-\omega} +\vert +\omega^{out}) a^{out}_{+\omega}
+ h.c.\right]\, ,\\
\vert -\omega^{out})& =&{1\over \sqrt{4\pi \omega}}  \left[
 \Theta(-v) \exp (i4M\omega \ln {-v\over
a})-\exp (-i\omega u)\right] \, , \nn \\
\vert +\omega^{out})& =&{1\over  \sqrt{4\pi
\omega}}\Theta(v) \exp (-i4M\omega \ln {v\over a})\, .\label{out}
\end{eqnarray}
Here $\vert - \o )$ has positive frequencies  on  ${\cal J}^+$ and vanishes at $r=0$ inside the shell\footnote{The first  term in the first equation Eq.(\ref{out}) is determined by the vanishing condition at $r=0$ inside the shell, defining the $u,v$ coordinates there  by requiring continuity of the metric across the shell.} 
and $\pm \o$ designates respectively out-states defined
on $\cal H$ and ${\cal J}^+$.

A convenient complete set of positive frequency wave functions  on ${\cal J}^-$ is  \begin{eqnarray}
\vert \pm\omega^{in}) &=&{\exp (\pm \omega 2\pi M)\over
 \sqrt{8\pi \omega\ \sinh (\omega 4\pi M)}}\, \Theta(v)    \exp (\mp
i4M\omega \ln {v\over a} )  \nn \\&+&{\exp (\mp \omega 2\pi M )\over
 \sqrt{8\pi \omega\ \sinh (\omega 4\pi M)}} \left[ \Theta(-v) \exp (\mp
i4M\omega \ln {-v\over a}) - \exp(\pm i\omega
u) \right]\, .\label{in} 
\end{eqnarray}
The -in and -out operators, $a_{\pm \o}^{in}$ and $a_{\pm \o}^{out}$,  are related by the Bogoljubov transformation 
 \begin{equation}
\label{bogo}
  a^{in}_{\pm\omega}
  =  \alpha_\omega a^{out}_{\pm\omega} - \beta_\omega a^{out
\dagger}_{\mp\omega} \qquad  \qquad \alpha_\omega =\frac {\exp
 (\omega 2\pi M)}{ \sqrt{ 2\sinh (\omega 4\pi M)}} \quad,\quad \beta_\omega
=\frac{\exp (- \omega 2\pi M)}{ \sqrt{ 2\sinh (\omega 4\pi
M)}}\, ,
\end{equation}
and the Heisenberg vacuum $\vert i\rangle$ is related to the out-vaccum  $\vert  \O\rangle $, where $\vert \O\rangle = \vert \O_{{\cal J}^+}\rangle \vert \O_{\cal H}\rangle $, by
 \begin{equation}
\label{vacuum}
\vert i \rangle =\langle\Omega \vert i \rangle \exp\left[\int
\frac{\beta_\omega}{\alpha_\omega}\, a^{out \dagger}_{-\omega} a^{out
\dagger}_{+\omega}\,d\omega\right]\vert \Omega  \rangle\, .
\end{equation}
The Hawking radiation obtains at the global temperature $ (8\pi M)^{-1}$,
  \begin{equation}
\label{hawking}
tr\Pi_{{\mathbf H}_ {\cal H}}\vert i\rangle \langle i\vert  =\frac{1}{Z}\exp{(-8\pi M})H\, ,
\end{equation}
where $H$ is the free field Hamiltonian and $Z$ its partition function.

The existence of the non-trivial Bogoljubov transformation Eq.(\ref{bogo}) follows from  Eq.(\ref{approx}). Thus, the Hawking radiation starts at a Schwarzschild time $t_0$ when the radius of the shell $r_s$ approaches the horizon within a coordinate distance $r_s-2M$ of order $M$. One introduces the ``proper distance" to the horizon 
 \begin{equation}
\label{proper}
\r (r) =\int_{2M}^r  g_{rr}^{1/2}(r^\prime) dr^\prime  \simeq (8M)^\frac{1}{2}(r-2M)^\frac{1}{2}  \, ,
\end{equation}  
 that measures the local Hawking temperature $T_{local} = 1/(8\pi M\sqrt{ g_{00}})\simeq 1/\r (r)$. Thus the local frequency of a Hawking quantum, measured on the star surface, reaches the Planck scale in a static frame  when $\r (r_s) = O(1)$, or equivalently  when $ r_s-2M = O(1/M)$. This happens at a time $t-t_0= O(4M\ln M)$.  As quanta with energy of $O(1/M)$ are emitted in a time $O(M)$, the evaporation of the black hole into $O(M^2)$ quanta takes a time of $O(M^3)$. This means that the evaporation requires transplanckian scales up to $\r ^{-1}=O(M^{-1}\exp \x M^2)$ with $\x =O(1)$ and that only a negligible fraction of the radiation occurs before the star reaches a Planckian vicinity of the horizon.  While these fluctuations do not affect significantly the metric of the inclusive background, they will play a crucial r\^ole  in  exclusive backgrounds, as will be seen in Section~3.3.3.

The above computation of the Hawking radiation Eq.(\ref{hawking}) illustrates in simple terms that unitarity is  verified in the space-time region limited by the horizon and that the thermal distribution arises solely from the trace over the states on the horizon. In fact, if one conceived the physically contrived experiment of bringing the classical shell at rest when $\r \ll O(1)$, the Hawking radiation would start as usual when the radius of the shell reaches $\r = O(M)$ and would stop when the shell reaches its final position. There would be no event horizon and the correlations between the quanta emitted as the shell is decelerating and of the earlier Hawking quanta would ensure unitarity. It may be useful to keep this thought experiment in mind when studying the back-reaction of the Hawking radiation in presence of a post-selection on $\cal J^+$ as it suggests a more genuine quantum way to connect unitarity with the fate of the event horizon.

\subsubsection{General decomposition of the weak energy momentum tensor}
Starting with the unperturbed background of Fig.1, we now  impose an out-state $\vert \p\rangle$ on ${\cal J}^+$. The Cauchy surface defining the out-states is ${\cal J}^+ \cup \cal H$ and the Hilbert space decomposes in the tensor product ${\mathbf H}={\mathbf H}_{{\cal J}^+}\otimes {\mathbf H}_ {\cal H}$. The expression of the   weak energy-momentum tensor for such a partial post-selection generalizes to  Eq.(\ref{partial}). It reads
\begin{equation}
\label{weakpartial}
  T_{\mu\nu[weak]}^{\,\p\, i\,(2)}   = \frac{\langle i\vert \p\rangle \langle \p\vert \hat T_{\m\n}^{(2)} \,\vert i \rangle}{\langle i \vert \p\rangle \langle \p \vert i\rangle}\, .
\end{equation}
The in-vacuum $\vert i\rangle$ is defined on ${\cal J}^-$ and $\langle  \p  \vert   i \rangle \in {\mathbf H}_ {\cal H}$ is the EPR conjugate of the state $\vert \p\rangle\in~{\mathbf H}_{{\cal J}^+}$.

The $\hat T_{uu}^{(2)}$ component of the energy-momentum tensor contains the relevant information about the Hawking radiation. We shall evaluate  $ T_{uu[weak]}^{\,\p\, i\,(2)}  $ near the horizon.   One has
\begin{equation}
\label{normal}
 \hat T_{uu}^{(2)} =\,  : \hat T_{uu}^{(2)}: + \langle i \vert \hat T_{uu}^{(2)}\vert i\rangle\, ,
\end{equation}
where
\begin{eqnarray}
\label{uunormal}
: \hat T_{uu}^{(2)}: \vert i\rangle= \lim_{u',u''\to u}\int\int d\omega d\omega^\prime \left[ \partial_{u'}\vert-\omega^{in}\,)^\star a^{in\dagger}_{-\omega}+ \partial_{u'}\vert+\omega^{in}\,)^\star a^{in\dagger}_{+\omega}\right]\nn\\\left[ \partial_{u''}\vert-\omega^{\prime\, in}\,)^\star a^{in\dagger}_{-\omega^\prime}+ \partial_{u''}\vert+\omega^{\prime\, in}\,)^\star a^{in\dagger}_{+\omega^\prime}\right]\vert i\rangle\, .
\end{eqnarray}
The Bogoljubov transformation Eq.(\ref{bogo})  implies
\begin{equation}
\label{inout}
  a^{in\dagger}_{\pm\omega}\, a^{in\dagger}_{\mp\omega^\prime}\vert i\rangle =\left[\frac{1}{\alpha _\omega \alpha_{\omega^\prime}} a^{out\dagger}_{\pm\omega}\,a^{out\dagger}_{\mp\omega^\prime}-\frac{\beta_\omega}{\alpha_\omega}\delta(\omega-\omega^\prime)\right]\vert i\rangle \, ,
 \end{equation}
 and we write
\begin{equation}
\label{outout}
: \hat T_{uu}^{(2)}:\vert i\rangle =\left\{:: \hat T_{uu}^{(2)}::   +  \frac{\langle\Omega\vert : \hat T_{uu}^{(2)}: \vert i\rangle}{\langle \Omega\vert i\rangle}\right\}\vert i\rangle ,
 \end{equation}
where $:: \hat T_{uu}^{(2)}::$ contains only pairs of out-state creation operators. Combining Eqs.(\ref{normal}) and  (\ref{outout}) we get~\cite{MP, Englert:1994qe}
\begin{equation}
\label{result}
 T_{uu[weak]}^{\,\p\, i\,(2)}   = \frac{\langle i\vert \p\rangle \langle \p\vert ::\hat T_{uu}^{(2)}:: \,\vert i \rangle}{\langle i \vert \p\rangle \langle \p \vert i\rangle}+ \frac{\langle\Omega\vert : \hat T_{uu}^{(2)}: \vert i\rangle}{\langle \Omega\vert i\rangle} +  \langle i \vert \hat T_{uu}^{(2)}\vert i\rangle\, .
 \end{equation}
One may interpret Eq.(\ref{result}) as follows: the last term in the r.h.s. is a renormalized expectation value which was evaluated in Eqs.(\ref{Tuu}), (\ref{tu}) and (\ref{TuuH}). 
The Hawking radiation ensures its regularity on the horizon, as seen in the scalar  $\langle T^2\rangle ^{(2)}_{Unruh}$,  Eq.(\ref{2hor}). The second term is the weak value of the final vacuum,  which thus contains no radiation. One expects that it will lead to  the divergence on the horizon characteristic of the static ``Boulware vacuum'', as will be checked in Section 3.3.3. Using Eqs.(\ref{Tvv}) and (\ref{Tuu}) with $t_v=t_u=0$, we get for the sum of the two last terms of Eq.(\ref{result}), in terms of the proper distance  to the horizon Eq.(\ref{proper}), 
\begin{equation}
\label{2Bhor}
\langle T^2\rangle ^{(2)}_{Boulware} \stackrel{r\to 2M}{\longrightarrow}\frac{1}{\r^4}\frac{1}{288\pi^2}\, .
\end{equation}
The divergence in Eq.(\ref{2Bhor}) represents the infinite {\it negative} vacuum energy of the Boulware vacuum on the horizon, as seen for instance in Eq.(\ref{TvvH}). As a consequence of the general sum rule Eq.(\ref{partialav}), this divergence cancels if one averages the weak values Eq.(\ref{result}) with probabilities $ \langle i \vert \p_k\rangle\langle \p_k\vert i\rangle$ over a complete orthonormal set $\vert \p_k \rangle$ on $\cal J^+$. One recovers in this way  the regular expectation value $\langle i \vert T_{uu}^{(2)}\vert i \rangle$, from which we constructed $\langle T^2\rangle ^{(2)}_{Unruh}$. One may view the first term in Eq.(\ref{result}) as a particular contribution to the radiation energy on top of a Boulware vacuum described  by the two last terms. 

We shall now discuss in more details the decomposition Eq.(\ref{result}) and extend our conclusions to the four dimensional case where Eq.(\ref{result}) similarly reads
\begin{equation}
\label{4result}
 T_{uu[weak]}^{\,\p\, i}  = \frac{\langle i\vert \p\rangle \langle \p\vert ::\hat T_{uu}:: \,\vert i \rangle}{\langle i \vert \p\rangle \langle \p \vert i\rangle}+ \frac{\langle\Omega\vert : \hat T_{uu}: \vert i\rangle}{\langle \Omega\vert i\rangle} +  \langle i \vert \hat T_{uu}\vert i\rangle\, .
 \end{equation}
This will allow us to obtain the linear back-reaction for partially post-selected states, as a preliminary step for formulating our conjectures about  S-matrix amplitudes in Section 4.

\subsubsection{Back-reaction of the partially post-selected Hawking radiation}

We first confirm the divergence Eq.(\ref{2Bhor}) of the vacuum terms. From Eqs(\ref{uunormal}), (\ref{inout}) and (\ref{bogo}) one gets
\begin{equation}
\label{boulware}
\frac{\langle\Omega\vert : \hat T_{uu}^{(2)}: \vert i\rangle}{\langle \Omega\vert i\rangle} = - \frac{1}{2\pi}\int_0^\infty d\omega \frac{\omega}{e^{8\pi M \omega}-1 }= \frac{-1}{(4M)^2}\frac{1}{48\pi}\, ,
\end{equation}
and hence,  using Eqs (\ref{Tuu}) and (\ref{tu}), we obtain in the vicinity of the horizon, for asymptotically large $v$,
\begin{equation}
\label{bvacuum}
 \frac{\langle\Omega\vert : \hat T_{uu}^{(2)}: \vert i\rangle}{\langle \Omega\vert i\rangle}+\langle i\vert  \hat T_{uu}^{(2)} \vert i\rangle ={1\over 12\pi} \left[-{M\over
2r^3}(1-{2M\over r}) - {M^2\over 4r^4}  \right] \, ,
\end{equation}
thus recovering  Eq.(\ref{Tuu}) with $t_u(u)=0$ and hence the divergent Boulware vacuum result Eq.(\ref{2Bhor})\footnote{The divergence sticks of course for finite $v$ (see footnote 3).}.

We now estimate the first term in the decomposition Eq.(\ref{result}), namely
\begin{equation}
\label{1term}
 T_{uu[weak]}^{\,\p\, i\,(2)_{\,rad}} \equiv \frac{\langle i\vert \p\rangle \langle \p\vert ::\hat T_{uu}^{(2)}:: \,\vert i \rangle}{\langle i \vert \p\rangle \langle \p \vert i\rangle}\, .
 \end{equation}
 We expand the state $\vert \p\rangle$ into a complete set of orthonormal states $\vert P_i\rangle $, which we take to be energy eigenstates. Thus 
\begin{equation}
\label{expand}
\vert\p\rangle = \sum_i \alpha_i \vert P_i\rangle  \quad\text{with}\quad  \vert P_i \rangle=\frac {1}{\sqrt{n_1!\,n_2!\dots n_n!}}(\tilde a_{-\omega_1}^{out \dagger})^{n_1}\dots(\tilde a_{-\omega_n}^{out \dagger})^{n_n}\vert  \O_{{\cal I}^+}\rangle\, .
\end{equation}
Here, we have discretized integrals over final states, using a cutoff time $\tilde\t$ for the emitted radiation, namely
\begin{equation}
\label{cutoff}
\sum_j = \frac {\tilde \t}{2\pi}\int d\omega \qquad \tilde a_{\omega_j} =  \left(\frac {2\pi}{\tilde\tau}\right)^{\frac{1}{2}}a_\omega\, .
\end{equation}
Letting $\tilde\t\to \infty$ leads to the conventional  description of the Hawking radiation in absence of back-reaction.  This limit is however unphysical as the total radiation energy on $\cal J^+$ would diverge, as it is only compensated by that of the partners independently  of the black hole mass. Introducing an energy conservation constraint would limit $\tilde\t$ to the evaporation time $O(M^3)$ but we shall here simply use $\tilde\t$ as a regulator in the unperturbed problem for computing energy  {\it densities}. From Eq.(\ref{vacuum}) and (\ref{expand}), one gets
\begin{equation}
\label{conjugate}
\langle P_i\vert i\rangle =\langle \O\vert i\rangle \frac {1}{\sqrt{n_1!\,n_2!\dots n_n!}}\exp \left(-4\pi M \sum_{k=1}^n \o_k^{(i)} n_k^{(i)}\right) (\tilde a_{-\omega_1}^{out \dagger})^{n_1}\dots(\tilde a_{-\omega_n}^{out \dagger})^{n_n}\vert  \O_{\cal H}\rangle\, .
\end{equation}

Eq.(\ref{conjugate}) shows that  the denominator $\langle i \vert \p\rangle \langle \p \vert i\rangle$ of $T_{uu}^{rad,\p\,(2)}$ is diagonal in the energy basis $\vert P_i\rangle$ and is given by
\begin{equation}
\label{denominator}
\langle i \vert \p\rangle \langle \p \vert i\rangle  =\langle i\vert \O\rangle \langle \O\vert i\rangle \sum_i \vert \a_i\vert^2 \exp \left(-8\pi M \sum_{k=1}^n \o_k^{(i)} n_k^{(i)}\right)\, .
\end{equation}
From Eq.(\ref{vacuum}), one has 
\begin{equation}
\label{partition}
[\langle i\vert \O\rangle \langle \O\vert i\rangle]^{-1} = \sum_i \exp (-8\pi M E_i) \qquad  E_i = \sum_{k=1}^n \o_k^{(i)} n_k^{(i)}\, ,
\end{equation}
and Eq.(\ref{denominator})  can be written as
\begin{equation}
\label{denominator1}
\langle i \vert \p\rangle \langle \p \vert i\rangle = \sum_i \vert \a_i\vert^2 p_i \, ,
\end{equation}
where $p_i$ is the thermal probability of finding the state $\vert P_i\rangle$ of energy $E_i$  at the temperature $ (8\pi M)^{-1}$.

The numerator  $\langle i\vert \p\rangle \langle \p\vert ::\hat T_{uu}^{(2)}:: \,\vert i \rangle$ of  $ T_{uu[weak]}^{\,\p\, i\,(2)_{\,rad}} $ contains both diagonal and non-diagonal terms. Eqs.(\ref{uunormal}), (\ref{inout}), (\ref{expand}), (\ref{conjugate}) and (\ref{partition}) yield for the diagonal contribution to the numerator of an energy state  $\vert P_i\rangle$ 
\begin{equation}
\label{numEi}
\langle i\vert  P_i\rangle \langle  P_i\vert ::\hat T_{uu}^{(2)}:: \,\vert i \rangle=\lim_{\tilde\t \to \infty}\frac {1}{\tilde\t}\, p_i E_i \, .
\end{equation}
Note that summing Eq.(\ref{numEi}) over the complete set of energy states and using the continuous limit Eq.(\ref{cutoff}), one checks
\begin{equation}
\label{complete}
\langle i\vert ::\hat T_{uu}^{(2)}:: \,\vert i \rangle=\frac{1}{2\pi}\int_0^\infty d\o\frac{\o}{e^{8\pi M\o}-1} =\frac{1}{(4M)^2}\frac{1}{48\pi}\, .
\end{equation}
This term cancels in the general partial post-selection sum-rule Eq.(\ref{partialav}) the contribution  of the weak value of the final vacuum $\vert\O\rangle$  in Eq.(\ref{result}) to recover the regular expectation value of $\hat T_{uu}^{(2)}$ given in Eq.(\ref{TuuH}).

We write the $\tilde\t$-regulated  $T_{uu[weak]}^{\,\p\, i\,(2)_{\,rad}} $ as the sum of its diagonal part  $\tilde\t$-regulated  $T_{uu[weak]}^{\,\p\, i\,(2)_{\,rad\, D}}$ and its non-diagonal one  $T_{uu[weak]}^{\,\p\, i\,(2)_{\,rad\, ND}}$. One has
\begin{equation}
\label{diagonal}
T_{uu[weak]}^{\,\p\, i\,(2)_{\,rad\, D}}\equiv \frac{\sum_i \vert \a_i\vert^2\langle i\vert P_i\rangle \langle P_i\vert ::\hat T_{uu}^{(2)}:: \,\vert i \rangle}{\sum_j \vert \a_j\vert^2\langle i \vert P_j\rangle \langle P_j \vert i\rangle}= \frac{1}{\tilde\t}\frac{\sum_i \vert \alpha_i\vert^2 p_i E_i}{\sum_i \vert \alpha_i\vert^2 p_i}\, ,
\end{equation}
while $T_{uu[weak]}^{\,\p\, i\,(2)_{\,rad\, ND}}$ exhibits oscillations on scale $\D u=O(M)$. These arise because the numerator of Eq.(\ref{1term}) contains complex terms $\a_k \a_l^\star \langle i\vert  P_k\rangle \langle  P_l\vert ::\hat T_{uu}^{(2)}:: \,\vert i \rangle \, \,  k\neq l$ when $ \vert  P_k\rangle $ and  $\vert  P_l\rangle$ differ by one unit in the 
number of photons for two frequencies $\o_a$ and $\o_b$. This induces a phase factor
$\exp  (\o_a -\o_b)u $ or $\exp \pm (\o_a +\o_b)u$, according to whether the total number of quanta in $ \vert  P_k\rangle $ and  $\vert  P_l\rangle$ are the same or differ by two units. For typical frequencies  $\o_a\simeq\o_b\simeq 1/M$, these phases give rise to  the aforementioned oscillations, allowing for the description of wave packets.  

Let us first consider the behavior of $T_{uu[weak]}^{\,\p\, i\,(2)_{\,rad\, D}}$ close to the horizon as experienced by a free-falling object. The r.h.s. of Eq.(\ref{diagonal}) will cancel the Boulware singularity on the horizon if and only if 
\begin{equation}
\label{tuned}
\lim_{\tilde \t \to\infty}\frac{1}{\tilde\t}\frac{\sum_i \vert \alpha_i\vert^2 p_i E_i}{\sum_i \vert \alpha_i\vert^2 p_i}=\lim_{\tilde \t \to\infty}\frac{1}{\tilde\t}\langle E \rangle\equiv\frac{1}{2\pi}\int_0^\infty d\o \frac{\o}{e^{8\pi M\o}-1} =\frac{1}{(4M)^2}\frac{1}{48\pi}\, .
\end{equation}
This would be the case if all the $\alpha_i$ were chosen to satisfy Eq.(\ref{tuned}), for instance, if all $\vert \alpha_i\vert$ were equal.\footnote{Such a tuned choice in the unperturbed geometry of Fig.3  might be more natural in a background consistent with back-reaction. It might mimic energy conservation in the evaporation process in a finite dimensional Hilbert space. This cannot be imposed in the unperturbed background where energy conservation is trivially satisfied (in an infinite dimensional Hilbert space) between the radiation energy and the negative energy crossing the horizon, independently of the star mass. Thus singularities on the horizon in Eq.(\ref{tuned}), and in Eq.(\ref{OT2}) hereafter, might be an artifact of the approximation. This should be kept in mind while carrying the following discussion in the framework of this approximate description.} Consider post-selections from states with comparable contributions to  $T_{uu}^{rad,\p ; D\,(2)}$ as the thermal average, namely $O(1/M^2)$. These we label {\it generic} post-selections. Close to the horizon, their weak energy density  $ T_{uu[weak]}^{\,\p\, i\,(2)}$ is, on the average, dominated by the sum of the Boulware energy  density Eq.(\ref{boulware}) and  $T_{uu}^{rad,\p ; D\,(2)}$. Thus
\begin{equation}
\label{natural}
 T_{uu[weak]}^{\,\p\, i\,(2)} \stackrel{r\to 2M}{\longrightarrow} -\frac{1}{(4M)^2}\frac{1}{48\pi} + O(\frac{1}{M^2})= O(\frac{1}{M^2})\, .
\end{equation}
This yields a singularity on the horizon in the free falling frame or in the scalar 
 $ \langle T^2\rangle ^{(2)}_{weak}$ defined by Eq.(\ref{T2}) with $T_{\m\n}^{(2)}$ now identified with the weak value $ T_{uu[weak]}^{\,\p\, i\,(2)} $. We evaluate $ \langle T^2\rangle ^{(2)}_{weak}$ with $ T_{uu[weak]}^{\,\p\, i\,(2)}$ given by Eq.(\ref{natural}),  $ T_{vv[weak]}^{\,\p\, i\,(2)}$ outside a Planckian vicinity of $v=0$ by the expectation value Eq.(\ref{TvvH}), and  the c-number $ T_{uv[weak]}^{\,\p\, i\,(2)}$  by the trace anomaly Eq.(\ref{TuvH}), to get the leading contribution close to the horizon:
\begin{equation}
\label{OT2}
 \langle T^2\rangle ^{(2)}_{weak} \stackrel{r\to 2M}{\longrightarrow}O\left(\frac{1}{\r(r)^4}\right) \, .
\end{equation}
This result characterizes the divergence of the weak scalar  $ \langle T^2\rangle ^{(2)}_{weak}$ for such generic post-selected states on $\cal J^+$, which is of course of the same order of magnitude as the divergence of the Boulware vacuum energy density given in Eq.(\ref{2Bhor}).  More generally, nearly all post-selections will have a divergence on the horizon.

Generic non-diagonal $T_{uu[weak]}^{\,\p\, i\,(2)_{\,rad\, ND}}$ have similar amplitudes\footnote{This is true whether or not the singularity implied by Eq.(\ref{natural}) is effectively realized (see footnote 6).}  but, as discussed above, oscillates on a scale $\vert\D u\vert \simeq 1/\o =O(M)$. Similarly $T_{vv[weak]}$ oscillates on the scale $\vert \D v\vert \simeq v/4M\o$ which is of order $O(M)$ in the vicinity of the star.  Close to the horizon we get
\begin{equation}
\label{fluc}
\D \langle T^2\rangle ^{(2)}_{weak}\simeq  \langle T^2\rangle ^{(2)}_{weak}\left[\frac{1}{M} \,(\vert\D u\vert +\vert\D v\vert+\vert\D r^\star\vert)\right]\, ,
\end{equation}
where the last term comes from the variation of $(g^{uv})^2$. Consider for instance a constant  $t$  line. The three terms are of order $M$ and hence $\D r =O(r-2M)$. When
the amplitude encoded in Eq.(\ref{OT2}) reaches the Planck scale, $\D r=O(1/M)$ or equivalently $\D\r(r) = O(1)$.  Thus  the amplitude and the spread of the oscillations of  $\langle T^2\rangle ^{(2)}_{weak;\p}$ reach simultaneously the Planck scale at an invariant Planck distance $\r =O(1)$ from the horizon for generic post-selected states. More generally, nearly all post-selected states will exhibit divergent oscillations in the vicinity of the horizon.

Up to now, we have considered $ T_{uu[weak]}^{\,\p\, i\,(2)}$ close to the horizon, outside the star. Inside the star, the Minkowskian metric can be written as
\begin{equation}
ds^2=  dU\  dV \qquad  \qquad   U= \t -r \quad ,\quad V= \t +rÊ\, ,
\label{Min2}
\end{equation}
and continuity of the metric close to the horizon across the light-like shell yields 
\begin{equation}
U= -4M \exp \frac{v_s - u}{4M}\quad ,\quad V= vÊ\, .
\label{continuous}
\end{equation}
For the generic post-selection considered, we get inside the star, near its surface, a value comparable to its value outside the star
\begin{eqnarray}
\label{inside}
 \langle T^2\rangle ^{(2)}_{weak} \stackrel{r\to 2M}{\longrightarrow} 8 T_{UU[weak]}^{(2)}\, T_{VV[weak]}^{(2)}& = & 8\left(\frac{\partial u}{\partial U}\right)^2 T_{uu[weak]}^{(2)}\, T_{vv[weak]}^{(2)}\nn\\&=& O\left(\frac{1}{\r(r_s)^4}\right) \, .
\end{eqnarray}

To get an idea on how  $ T_{uu[weak]}^{\,\p\, i}$ could affect the geometry in the Planckian vicinity of the horizon, it is mandatory to include, at least at a qualitative level, the contribution to the weak energy-momentum tensor of the higher angular momentum modes  in the genuine four-dimensional case. First we notice that for the spherically symmetric s-wave contributions to generic weak values, we may approximately relate, as in Section 3.2, $T_{\m\n[weak]}^{\,\p\, i\,(2)}$ to its four-dimensional counterpart $T_{\m\n[weak]}^{\,\p\, i}$ by $T_{\m\n[weak]}^{\,\p\, i\,(2)}=4\pi r^2\, T_{\m\n[weak]}^{\,\p\, i}$. Thus for the s-wave contributions, we get
\begin{equation}
\label{sweak}
 \langle T^2\rangle_{weak; s} \stackrel{r\to 2M}{\longrightarrow}O\left(\frac{1}{M^4\r(r)^4}\right) \, .
\end{equation}
More generally, we may estimate in the vicinity of the horizon the contribution to generic weak values of higher angular momenta by computing the radiation from the full radial equation instead of the two-dimensional one, Eq.(\ref{phi}). This reads
\begin{equation}
\label{radial4}
\left(\frac{\partial^2}{\partial t^2}-\frac{\partial^2}{\partial r^{*2}}\right)\f + \left(1-\frac{2M}{r}\right)\left(\frac{2M}{r^3} +\frac{l(l+1)}{r^2}\right) \f =0 \, .
\end{equation}
We see that, while s-waves can easily escape to infinity, providing the dominant contribution to the Hawking radiation at infinity, high angular momenta contributions have small tunneling through the barrier. Their ancestor form a thermal atmosphere in the region between the horizon and $r<3M$. For generic weak values, we may estimate $ T^0_{0[weak]}$ by its thermodynamic average, {\it i.e.} 
\begin{equation}
\label{thermo}
T^0_{0[weak]}(r)\simeq\left(\frac{1}{8\pi M \sqrt{g_{00}(r)}}\right)^4 = O\left( \frac{1}{\r(r)^4}\right)\, .
\end{equation}
Other components are of the same order of magnitude, or vanishing.  We thus obtain\footnote{Notice that this result is related to the number $N(r)$ of modes which can cross $\r(r)$ before being reflected. To reach a radius $r$, they must have a Schwarzschild energy $\o\simeq 1/M $ such that
\begin{equation*}
\o^2>\left(1-\frac{2M}{r}\right)\left(\frac{2M}{r^3} +\frac{l(l+1)}{r^2}\right)  \, ,
\end{equation*}
and thus 
\begin{equation*}
l^2\lesssim\frac{M^2}{\r(r)^2}\, .
\end{equation*}
 Thus the number of modes $N(r)$ which can cross $\r(r)$ before being reflected is
\begin{equation*}
N(r)\simeq \frac{M^2}{\r(r)^2}\, .
\end{equation*}
The particles crossing $\r(r)$ thermalize and the contribution of each mode to the generic weak value of the energy density $T^0_{0[weak]}$ in the vicinity of the horizon  is roughly the same. This also applies to the escaping s-wave quanta because they spend roughly the same Schwarzschild time $O(M)$ before reaching the potential barrier as higher angular momentum modes do before being reabsorbed by the horizon; this yields an average static distribution because the emission rate is $O(1/M)$~\cite{Casher:1996ct}. From Eq.(\ref{sweak})  and the value of $N(r)$ we recover the estimate  Eq.(\ref{4weak})
\begin{equation*}
\langle T^2\rangle_{weak}  \,  \stackrel{r\to 2M}{\longrightarrow}O\left(\frac{1}{M^4\r(r)^4}\right). \left(\frac{M^2}{\r(r)^2}\right)^2 = O\left(\frac{1}{\r(r)^8}\right)\, .
\end{equation*}}
\begin{equation}
\label{4weak}
 \langle T^2\rangle_{weak}  \,  \stackrel{r\to 2M}{\longrightarrow} O\left(\frac{1}{\r(r)^8}\right)\, .
\end{equation}
The sum rule Eq.(\ref{partialav}) and the decomposition Eq.(\ref{4result}) indicates that the divergent generic weak value Eq.(\ref{4weak}) should be of the same order as the the Boulware vacuum energy in the vicinity of the horizon. This value for  $ \langle T^2\rangle_{Boulware} $ is in agreement with  the result of reference~\cite{Candelas}.
Eq.(\ref{steep}) yields now for the linear back-reaction of the generic partially post-selected amplitude 
\begin{equation}
\label{backweak}
R_{\m\n}^0 R^{\m\n \,0} = 64 \pi^2 \langle T^2\rangle_{weak}\, .
\end{equation}
with the weak value on the r.h.s. evaluated  in the Schwarzschild background $g_{\m\n}^{\bar 0}$. As a consequence of the estimate Eq.(\ref{4weak}),  the linear approximation, and presumably also the field theoretic description of the metric, should break down around $\r(r)= O(1)$ for partially post-selected amplitudes. This should be true due to the oscillations in  $\langle T^2\rangle_{weak}$, even if the  singularities on the horizon in weak values were disregarded (see footnotes 4 and 5).

The classical  background arising from the saddle point contribution to  generic amplitudes post selected on $\cal J^+$ appears to break down at an invariant Planck distance from the horizon where the dominant contribution to the Hawking radiation occurs. One might have objected that the transplanckian fluctuations of the weak energy-momentum tensor in the vicinity of the horizon are irrelevant as compensation is expected to occur when an incoming object feels transplanckian effects on {\it both} sides of the horizon~\cite{giddings}. But, independently of the fact that the present analysis does not require information from inside the horizon, the fluctuations are inscribed in the geometry itself as seen from Eq.(\ref{backweak}), leading to inconsistency. We take the inadequacy of the classical geometry to describe such huge fluctuations as an indication that generic post-selection destroys the event horizon and that, for such amplitudes, classical geometry must be abandoned  in favor of a genuine quantum description, which unfortunately is at present not available operationally.

In view of the expected dramatic effect of post-selection on the fate of the event horizon, we have to reconsider the outcome of the full black hole evaporation as qualitatively depicted for instance in Fig.2b. This will be discussed in the forthcoming Section 4, but clearly, as no reliable model for full black hole evaporation does hitherto exist, our  conclusions will be conjectural. The features arising from post-selection do not appear in the inclusive background defined in absence of post-selection. In that case,  the background  driven by $\langle i \vert \hat T_{\m\n}\vert i\rangle$ is totally insensitive to Planckian gravitational effects. It does not, for time small compared to the black hole lifetime, depart significantly from the classical background geometry with its event horizon. How this can be reconciled with the effect of post-selection will now be examined.

\section{Unitary S-matrix and classical physics : the conjecture }

As pointed out in Section~2, we limit our space-time description to neighborhoods of saddle-points of a functional integral given in terms of local fields, but whose expression in terms of possible more fundamental quantities is hitherto unknown. These saddle-points geometries we labelled as backgrounds.

Taking as tentative background for the evaporating black hole the Penrose diagram depicted in Fig.2b, a post-selection on $\cal J^+$ would generate inconsistencies in the vicinity of the horizon, as previously. Planckian geometry fluctuations still are triggered by the quantum radiation. Following the conclusions of Section 3.3.3, we thus assume that in the full quantum history of the incipient black hole the horizon does not survive post-selection on $\cal J^+$. The description of an exclusive S-matrix amplitude should then be depicted on a asymptotically flat background void of horizon and hence of a future spatial singularity. Thus, the partial post-selection on $\cal J^+$, as performed in Section~3, becomes a full post-selection.  As inclusive amplitudes are coherent superpositions of sets of exclusive ones defined by basis of final states, no classical horizons is needed in the exact quantum description. The absence of classical horizons  leaves no compelling reason to dispute S-matrix unitarity~\cite{'tHooft:1984re}. 

\begin{figure}[h]
\centering
  \includegraphics[width=4cm]{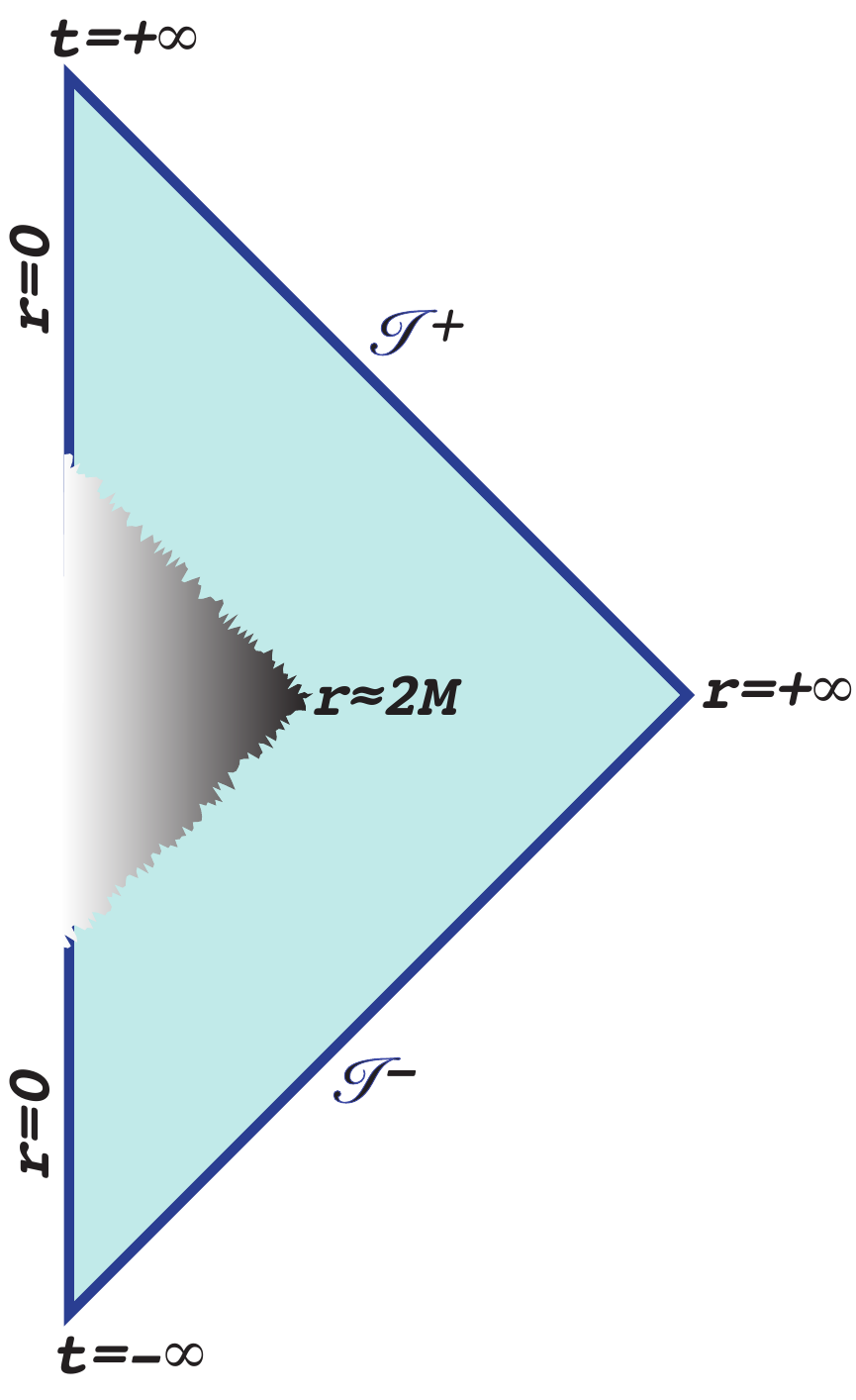} 
 \caption { \sl \small Penrose diagram for a black-hole S-matrix background. The shaded area extrapolating between two classical $r=0$ segments is dominated by quantum fluctuations} 
\end{figure}

The classical background of an S-matrix exclusive amplitude for a collapsing star of mass $M$ is thus confined outside the region of the classical horizon  to the region $\r \gtrsim1$. The S-matrix is assumed to be TCP invariant, the usual evaporation through Hawking radiation being favored only because of its very large relative phase space~\cite{Stephens:1993an}. A typical  amplitude (not necessarily TCP invariant), free of horizon and space-like singularity is sketched in Fig.~4. It is inspired by reference~\cite{Stephens:1993an} and comments by  't Hooft \cite{ws}. The classical region is depicted in blue and spans a region outside  the would-be horizon which first increases from $r=0$ to $r=2M$, then decreases back to $r=0$, mimicking the black hole birth and death. The shaded region of the drawing extrapolating between two classical $r=0$ segments, is assumed generically to be dominated by quantum fluctuations and might not have a well-defined space-time representation. 

How does one recovers the classical horizon and more generally the motion of classical objects in space-time?

A background with horizon is expected to appear for some suitable in-state  in the inclusive amplitude, when  the quantum superposition of exclusive backgrounds is disregarded in favor of a single saddle. No post-selection is performed and this inclusive background selects  in Table~1 the expectation value of the energy-momentum tensor as the quantum source of the Einstein equation Eq.(\ref{steep}). Such background appears indeed to be endowed  with an event horizon and to be free of transplanckian fluctuations~\cite{PP}. In view of this insensitivity to quantum gravity effects, we assume that it provides the classical approximation to the quantum black hole. Its conventional  Penrose representation, sketched in Fig.2b,  can easily be accommodated to represent  a real collapsing star. This approximation results in an apparent loss of unitarity. 

Let us first discuss the nature of the approximation outside of the horizon. There, the inclusive background is the statistical average Eq.(\ref{average}) of the exclusive backgrounds\footnote{In Section 2 the star was considered as a classical object in Eq.(\ref{steep}). It should now be included in the set of quantum fields, namely in the notation of Eq.(\ref{amplitude}), Eq.(\ref{steep})   now reads $ E_{\m\n}^0\equiv R_{\m\n}^0-\frac{1}{2} g_{\m\n}^0 R^0  = 8\pi \langle f\vert \hat T_{\m\n}^{\{\f_j\}}\vert i\rangle_0/\langle f\vert  i\rangle_0$.}.
The latter are approximately equivalent to the inclusive one only outside the Planckian region $\r(r)\simeq 1$, where there is barely radiation. We recall that the averaging is a consequence of the denominator appearing in the weak values driving the backgrounds. The classical event horizon thus emerges from the quantum interferences as a statistical average over horizonless geometries. As a consequence, the Bekenstein-Hawking horizon entropy ${\mathbf S} =4\pi M^2$ must be interpreted as a coarse-grained entropy.  Consistency requires that the classical geometry, endowed with an event horizon, and in fact the classical world, has only coarse-grained significance. It is on such an average geometry, depicted in blue in Fig.1, that the structureless Hawking radiation  was computed and appears as a valid approximation. 

Inside its horizon the average is ill-defined. However the inclusive background driven by the average energy-momentum tensor does extend in that region, which for exclusive amplitudes was assumed to be essentially quantum in character. We take this enhanced discrepancy between classical and quantum theory as a further confirmation of  the coarse-grained nature of the horizon and of the black hole classical geometry. Upon such background, the  quantum description  of matter is limited by its graininess and there is no contradiction with unitarity. 

The coarse-grained entropy ${\mathbf S}$ is equal to $\ln N$, where $N$ is the dimension of the Hilbert space. For a particular black hole, $N$ is the number of exclusive amplitudes defined by a complete set of out-states. The above description of the black hole geometry is reminiscent of the tentative ``fuzzball" description of static extremal black holes~\cite{MST}. Here  the fuzzball histories appear in exclusive processes  and the r\^ole of microstates is played by the complete set of the $N$ out-states. While our description of fuzzballs is more general and essentially dynamical in character, it clearly lacks the detailed description of the static fuzzball and its tentative interpretation in the AdS/CFT correspondence. 

The interpretation of horizons and more generally of classical geometries as a coarse-grained structure with thermodynamical significance is not new. Quite apart of its possible relation to fuzzballs, the analogy of the integrated Einstein equations with the law of thermodynamics was discovered by Bardeen, Carter and Hawking~\cite{BCH}. A local version was proposed by Jacobson~\cite{Jacobson} and generalized by many authors~\cite{local}. It was even argued that the thermodynamical significance of classical general relativity originates in the holographic principle~\cite{Verlinde}. 

To conclude, let us summarize our conjecture. We tentatively relate the exact (hitherto unknown) quantum description of Schwarzschild black holes to a classical description of the collapse. The quantum evolution is encoded in a unitary S-matrix constructed out of  complete sets of exclusive amplitudes; these would admit backgrounds void of horizon and space-like singularities, as proposed by 't Hooft. The classical approximation results from an approximate description of the inclusive  amplitude by the inclusive background, which disregards the quantum superposition of exclusive ones. One recovers in this way the classical picture endowed with an event horizon as a coarse-grained structure, extracted  from a unitary quantum-mechanical S-matrix for which no such horizon is needed.

While the computation of the full exclusive amplitudes would require a detailed operational theory of quantum gravity, the above scheme itself, which appeals to the metric description only in the vicinity of saddle-points, does not. However it does imply that unitary black hole possess in the classical approximation coarse-grained horizons and geometries.  Assessing  our conjecture requires more insight about horizon structure in general, and thus of its implications for cosmological and Rindler horizons. Such considerations are differed to a separate investigation.

\section*{Acknowledgments}
This work was supported in part by IISN-Belgium (conventions 
4.4511.06 and  4.4505.86), and by 
the Belgian Federal Science Policy Office through the 
Interuniversity Attraction Pole P~VI/11.

We are greatly indebted to Serge Massar for important comments covering all parts of this work. Fran\c cois Englert benefitted from his participation at early stages of this work to the 2008 CERN TH Institute  {\it``Black holes: a Landscape of theoretical physics problems''} organized by L. Alvarez-Gaume, N. Brambilla, C. Gomez, S. Ferrara and E. Rabinovici, and in particular from conversations with  Marika Taylor and Kostas Skenderis who  suggested the relation of our approach with the fuzzball conjecture.  He also  thanks Lenny Susskind for an illuminating discussion.  Philippe Spindel reiterates his thanks to IHES, where part of this work was elaborated.

\end{document}